\documentclass[amsmath, amssymb, prb, twocolumn, superscriptaddress, longbibliography, hypertext, floatfix]{revtex4-2}
\usepackage[utf8]{inputenc}
\usepackage[english]{babel}
\usepackage[T1]{fontenc}
\usepackage{graphicx}
\usepackage{braket}
\usepackage[usenames]{color}
\usepackage{hyperref} 
\hypersetup{colorlinks=true, linkcolor=blue, filecolor=magenta, urlcolor=cyan}

\newcommand{\Figref}[1]{Fig.~\ref{#1}}
\newcommand{\Eqref}[1]{Eq.~(\ref{#1})}
\newcommand{\Secref}[1]{Sec.~\ref{#1}}

\newcommand{\SM}[1]{SM Sec.~#1 \cite{supplementary}}
\newcommand{\ketbra}[2]{|{#1}\rangle\!\langle{#2}|}
\newcommand{\tr}[1]{\textup{Tr}\left(#1\right)}
\newcommand{\op}[1]{#1}
\newcommand{\opd}[1]{{#1}^\dagger}
\newcommand{\andres}[1]{\textcolor{cyan}{\bf #1}}
\newcommand{\dipc}{Donostia International Physics Center (DIPC), E-20018 Donostia-San Sebastián, Spain}
\newcommand{\loma}{Univ.~Bordeaux, CNRS, LOMA, UMR 5798, F-33405 Talence, France}

\begin{document}

\title{Single-molecule electroluminescence: crossover from weak to strong coupling}

\author{Andrés Bejarano}
\affiliation{\dipc}
\affiliation{Universidad del País Vasco (UPV/EHU), E-20018 Donostia-San Sebastián, Spain}
\affiliation{\loma}
\author{Moritz Frankerl}
\affiliation{\dipc}
\author{Rémi Avriller}
\affiliation{\loma}
\author{Thomas Frederiksen}
\affiliation{\dipc}
\affiliation{Ikerbasque, Basque Foundation for Science, E-48013 Bilbao, Spain}
\author{Fabio Pistolesi}
\affiliation{\loma}

\date{\today}

\begin{abstract}
We develop a microscopic model to investigate current-induced light emission in single-molecule tunnel junctions, where a two-level system interacts with a plasmonic field. Using the quantum master equation, we explore the transition from weak to strong plasmon-molecule coupling, identifying three distinct regimes governed by cooperativity, which quantifies the interplay between interaction strength and losses. Our findings establish a framework to detect strong coupling, unveiling resonance-dependent features in the emission spectrum and photon correlations.
\end{abstract}

\maketitle
Understanding and controlling light emission at the nanoscale is a fundamental challenge in modern nanophotonics and quantum optics \cite{ObFuVu.09.Photonicquantumtechnologies}. 
Among the various experimental platforms, the scanning tunneling microscope (STM) has proven to be a versatile tool for probing molecular-scale light-matter interactions with unparalleled spatial and spectral resolution.
Indeed, STM-induced electroluminescence, driven by inelastic tunneling electrons, has been extensively reported, particularly in plasmon-mediated light emission \cite{qiu2003, zhang-r2013, reecht2014, chong2016, zhang-li2017, imada2017, zhang-yao2017, chong2018, doppagne2018, kroger2018, chen2019, farrukh2021, vasilev2022, hung2023, dolezal2024, kaiser24prl}.  
A notable characteristic of these systems is the consistent observation of antibunching behavior in the second-order photon correlation function, indicating a low probability of emitting two photons simultaneously.
This behavior is a clear signature that the system operates as a single-photon source \cite{merino2015, zhang-li2017, roslawska2020, kaiser2025}.
Achieving strong coupling between molecular electronic states and the plasmonic mode could unlock exciting new possibilities for quantum manipulation, mirroring breakthroughs in atomic physics  \cite{haroche2001}. 
In particular, it could lead to the observation of Rabi oscillations between photonic and electronic modes, opening the door to novel quantum phenomena.
These desirable properties are particularly relevant for the advancement of quantum technologies, including quantum computing \cite{couteau2023}, quantum cryptography \cite{bozzio2022}, and precision sensing \cite{hadfield2009}.

The realization of strong coupling in single-molecule STM junctions remains a significant challenge, despite reports of Rabi-like double peak structures in the emission spectra from plasmonic junctions \cite{chikkaraddy2016, liu2017, paoletta2024}. 
Thus, most theories describing plasmon-molecule interactions assume weak coupling.
The decay of electronically excited states emits light into a structured photonic bath, representing the strongly damped plasmonic mode \cite{garraway1997, petruccione2002, neuman2018, jiangprl2023, jiangsc2023, zhang-yao2020, kaiser2025}. 
These theories cannot describe Rabi physics.
While the properties of light emission in tunnel junctions have been studied theoretically \cite{galperin2005, galperin2006, miwa2013, miwa2014, miwa2015, wang2018, vandenberg2019, miwa2019, lei2019}, a unified theoretical framework that captures both weak and strong coupling regimes—including hallmarks of strong coupling such as Rabi oscillations—is still lacking.
Identifying Rabi oscillations solely from the evolution of a double-peak structure in the spectrum is challenging in experiments, as tuning the system's physical parameters is difficult.
%
%
To clearly identify the coupling regime in current and future experiments, it is then crucial to develop a theory that accounts for both weak and strong coupling to predict experimentally accessible quantities.

Here, we present a theoretical framework based on the quantum master equation that spans from weak to strong coupling. This extends our previous work on the single-level model \cite{schaeverbeke2019, avriller2021}.
For clarity, we consider a system where the molecule is described as a two-level system (TLS) and the plasmonic cavity as a single damped electromagnetic mode.
The TLS can be charged or discharged through a typical electron tunneling rate of $\widetilde\Gamma$,
as schematically depicted in \Figref{fig:1}(a). 
The plasmonic cavity is coupled to the electromagnetic environment resulting in a damping rate $\kappa$.  
Surprisingly, we identify three distinct regimes determined not only by the ratio of the plasmon–molecule coupling strength $\Lambda$ to the damping rate $\kappa$, but also by the cooperativity parameter $C = \Lambda^2 /\widetilde\Gamma\kappa$, which plays a role analogous to that in quantum optics \cite{kimble1998, aspelmeyer2014}.
Specifically, we find that, in the weak coupling regime ($\Lambda\ll \kappa$), large or small cooperativities lead to parametrically different spectral widths of the molecular peak and a different short-time dependence of the second-order correlation function.
These differences in the correlation functions appear on the time scale fixed by the experimentally accessible inverse of the electronic tunneling time.
By contrast of the Rabi oscillations for strong coupling, that appear only on a much shorter scale, given by the inverse of the plasmonic lifetime, which could be challenging to observe.
Therefore, our results provide a detailed framework for understanding the interplay between coupling strength, cooperativity, and photon statistics in STM-based light emission experiments.


\begin{figure}[ht] 
    \includegraphics[width=\columnwidth]{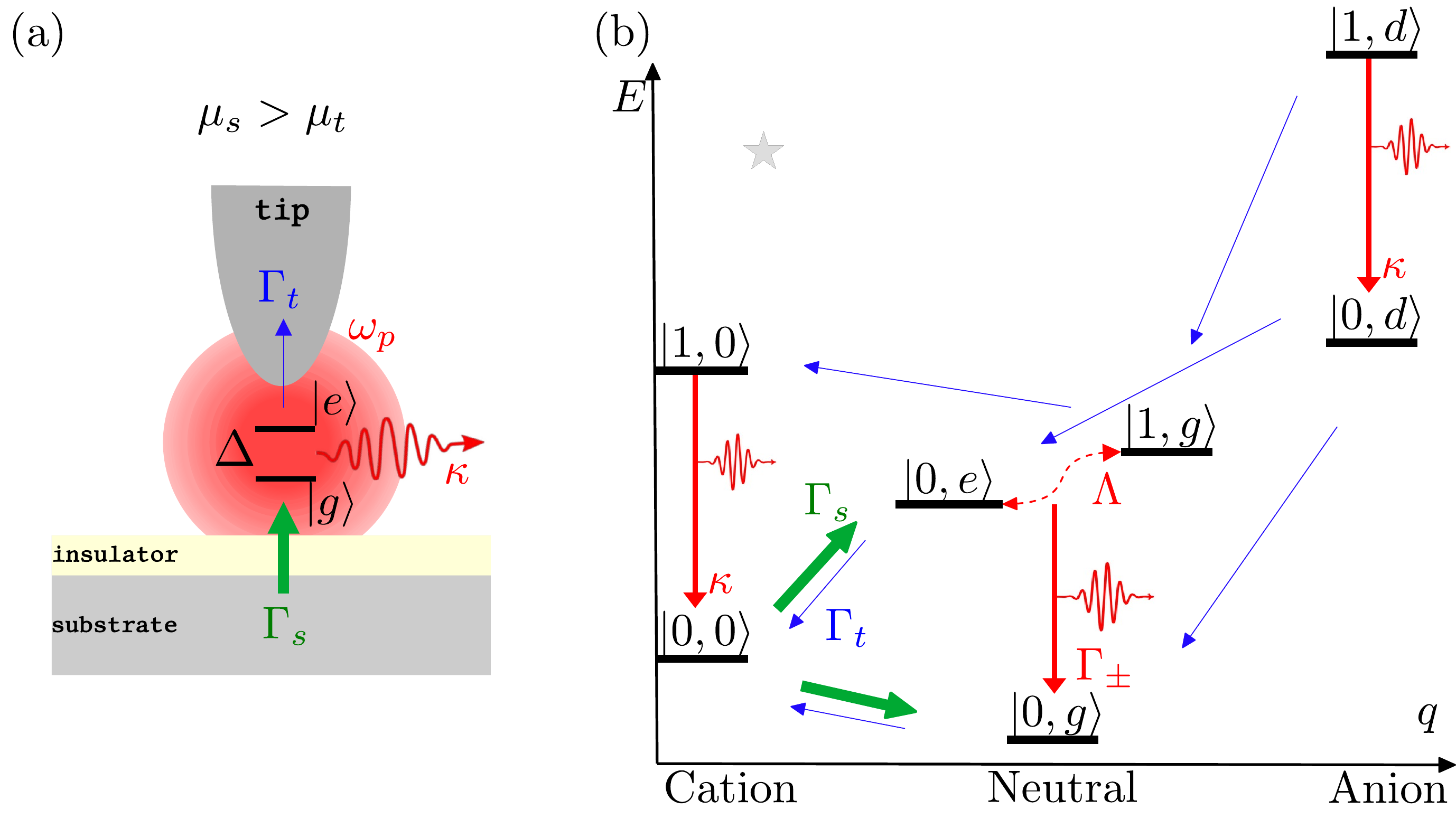}
    \includegraphics[width=\columnwidth]{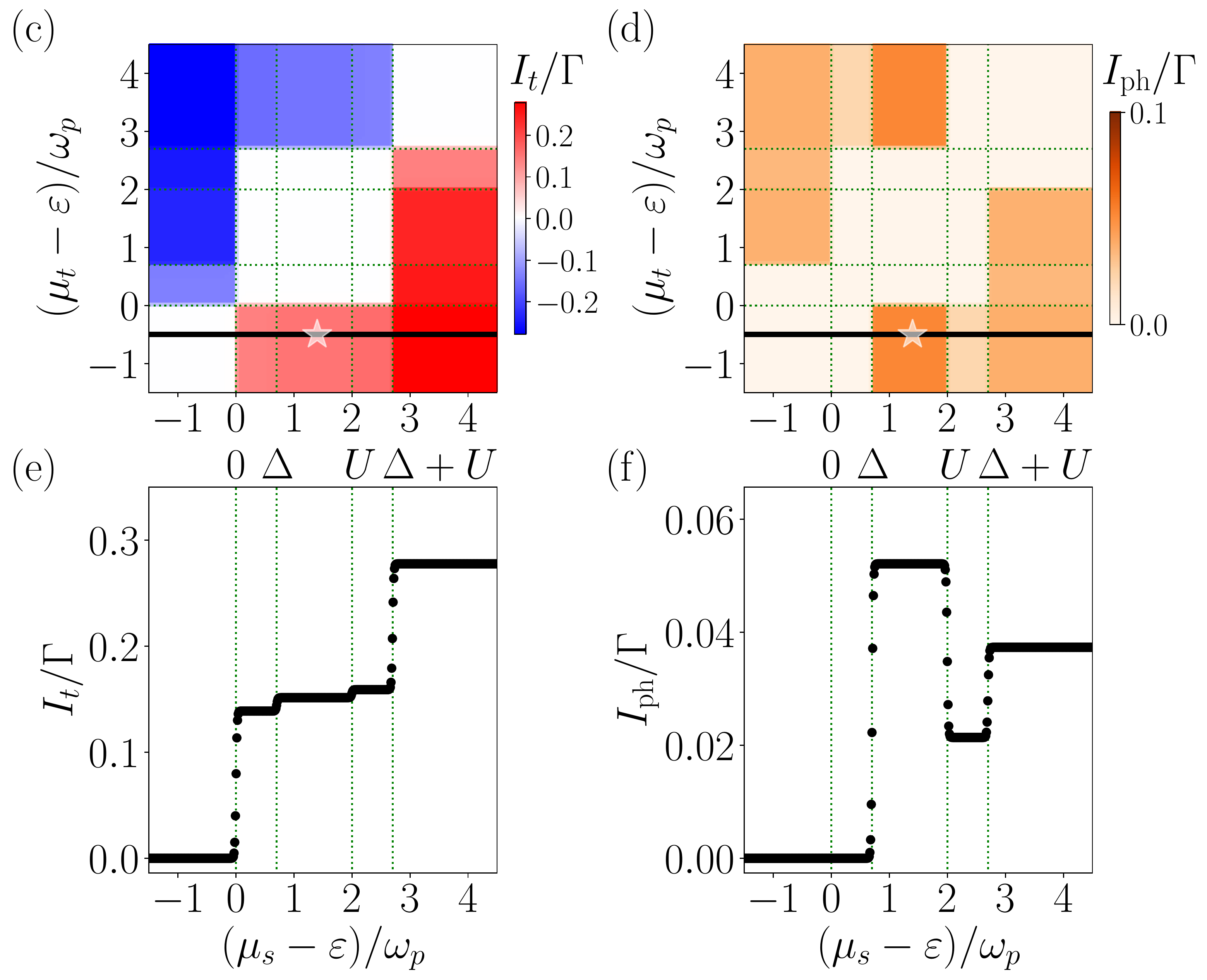}
    \caption{Model of single-molecule electroluminescence.
    (a) Schematic of a two-level electronic system coupled to electrodes and a plasmonic mode.
    (b) Many-body state diagram showing possible transitions within the plateau defined by $\Delta < \mu_s - \varepsilon < U$ and $\mu_t - \varepsilon < 0$. Within this region, we consider the bias configuration $(\mu_s, \mu_t) = \varepsilon + (1.4, -0.5)\omega_p$, marked by a $\star$ in panels c–d.
    Green (blue) arrows represent the possible transitions, at this specific bias configuration, corresponding to electron addition (extraction) via tunneling involving the substrate (tip) electrode.
    Similarly, red arrows represent transitions involving photon emission from the system. 
    (c,d) Electron ($I_t$) and photon ($I_{\mathrm{ph}}$) currents as functions of the shifted chemical potentials, scaled by $\Gamma=\Gamma_s+\Gamma_t$.
    (e,f) Slices of the currents as a function of $\mu_s$ at fixed $\mu_t-\varepsilon=-0.5\omega_p$ (black line in c-d).
    The calculations correspond to $\Lambda/\kappa = 0.04$, $\Delta = 0.7\omega_p$, $\varepsilon = -0.4\omega_p$, $U = 2\omega_p$, $\kappa = 0.05\omega_p$, $k_BT = 0.01\omega_p$, and $\Gamma_s=5\Gamma_t = 5\times 10^{-6}\omega_p$.}
    \label{fig:1}
\end{figure}
\textit{Model and approach}---In order to investigate the system depicted in \Figref{fig:1}(a) we introduce the following Hamiltonian (with $\hbar=1$):
\begin{equation}
\begin{aligned}
\op{H}_{\rm S} = &~  \varepsilon \opd{d}_g \op{d}_g+(\varepsilon +\Delta)  \opd{d}_e \op{d}_e+ U \opd{d}_g \opd{d}_e  \op{d}_e \op{d}_g 
\\
&~ +
\omega_p \opd{a} a+ 
\Lambda(\opd{a} \opd{d}_g \op{d}_e+a \opd{d}_e \op{d}_g) .
\label{eq:Hs}
\end{aligned}
\end{equation}
Here, the molecular two-level electronic system with energy gap $\Delta$, onsite Coulomb repulsion energy $U$, and single-level energy $\varepsilon$ is described by the first line in terms of Fermi annihilation operators $d_\sigma$ ($\sigma= g,e$) for the ground and excited state, respectively.
These single-particle states represent, for instance, the highest occupied and lowest unoccupied molecular orbitals (HOMO/LUMO) of a molecule that is placed between a substrate and an STM tip.
The plasmon forms between the tip and the substrate. It is described as a single cavity mode with resonance frequency $\omega_p$ and annihilation operator $a$.
The TLS interacts with this plasmon via dipolar coupling of strength $\Lambda$, within the rotating wave approximation, valid for $  \Lambda\ll \Delta, \omega_p $.
The purely electronic part of the Hamiltonian is diagonal and we can label the four manybody eigenstates with the occupancy of the levels as $|q\rangle$, with $q \in \{0,g,e,d\}$, where $|0\rangle$ is the vacuum, $|\sigma\rangle=d_\sigma^\dagger |0\rangle$, and $|d\rangle=d^\dagger_{g} d^\dagger_e |0\rangle$  the doubly occupied state. 
The eigenenergies read $\epsilon_q = 0$, $\varepsilon$, $\varepsilon+\Delta$, $2\varepsilon+\Delta+U$, respectively.

The molecular system is coupled to metallic leads ($\alpha=t,s$), allowing the tunneling of electrons with rates $\Gamma_{\alpha\sigma}$. Also, the plasmonic mode is coupled to an electromagnetic environment enabling emission of photons with a rate $\kappa$.
These system-bath interactions are further discussed in Sec.~S1.A of the Supplementary Material (SM) \cite{supplementary}.
The typical range of parameters in STM experiments is 
$\Gamma_{\alpha\sigma} \ll \kappa  \sim k_B T \ll \omega_p$,
where $T$ is the temperature and $k_B$ the Boltzmann constant.
These conditions allow us to use a Born-Markov approach to describe the coupling to the environment (see \SM{S1.B}).
Tracing out the environment degrees of freedom leads to a linear equation for the reduced density matrix $\rho$ of the system: $\dot \rho={\cal L} \rho$, with the Liouvillian 
$\mathcal{L} = \mathcal{L}_{\rm c} + \mathcal{L}_{\rm e}^+ +\mathcal{L}_{\rm e}^- + \mathcal{L}_{\rm ph}$.
The first term $\mathcal{L}_{\rm c}\rho=-i[H_{\rm S}, \rho]$ represents the coherent evolution of the system. 
The second and third terms
$
\mathcal{L}_{\rm e}^\pm=\sum_{\alpha\sigma qq'}
\Gamma_{\alpha\sigma}f_\alpha^\pm(\epsilon_{qq'})\mathcal{D}\big[(D^\pm_\sigma)_{q'}^{q}\big]$,
correspond to electron tunneling dissipators, which describe the addition or removal of electrons from the system.
Here we defined 
${\cal D}[X]\rho=X\rho X^\dagger -\{X^\dagger X,\rho\}/2$,
$\epsilon_{qq'}=\epsilon_q-\epsilon_{q'}$,
$D_\sigma^+=d_\sigma^\dagger$, $D_\sigma^-=d_\sigma$, 
$(D)^{q}_{q'}=\sum_n\ketbra{n,q}{n, q} D \ketbra{n,q'}{n, q'}$, 
$f^\pm_\alpha(\epsilon)=1/(e^{\pm(\epsilon-\mu_\alpha)/k_B T}+1)$, and $\mu_\alpha$ is the chemical potential of lead $\alpha$.
The last one is the photon dissipator for $k_B T \ll\omega_p $:
$\mathcal{L}_{\mathrm{ph}}=\kappa {\cal D}[a]$.
Following the literature of quantum optics \cite{cohen-tannoudji_atom_1998}, the three dissipators have been derived in the independent velocity approximation, {\em i.e.}, by neglecting the interaction term $\Lambda$. 
This formulation yields a Lindblad form, the details are provided in \SM{S1.B}.

The present approach enables the study of photon emission and electronic transport in the system, covering a wide range of coupling strengths, from the weak-coupling regime ($\Lambda \!\ll\! \kappa$) to strong coupling 
($\omega_p\!\gg\!\Lambda\!\gg\! \kappa$).
The average electronic current ($I_\alpha$) and photon current ($I_{\text{ph}}$) are obtained by averaging the jump operators defined by the dissipators entering the Born-Markov master equations \cite{marcos2010}.
Specifically, the jump operators adding/removing electrons read
${\cal J}_{e,\alpha}^\pm \rho 
=\sum_{\sigma qq'}\Gamma_{\alpha \sigma}f_\alpha^\pm(\epsilon_{qq'}) (D_\sigma^\pm)^{q'}_{q}\rho (D_\sigma^\mp)^{q}_{q'}$,
while the photon jump operator is ${\cal J}_{\rm ph} \rho=\kappa a\rho a^\dagger$
(see \SM{S1.E}).
We use the quantum regression theorem to obtain the photon spectrum 
$S(\omega)
=
{\kappa} {\rm Re} 
\int_{-\infty}^\infty d\tau {\rm Tr}
(a^\dagger e^{{\cal L}\tau}a \rho^{\textup{st}})
/2\pi$ and the second-order correlation function
$g^{(2)}(\tau\geq 0)={\rm Tr} (a^\dagger a e^{{\cal L}\tau}[ a \rho^{\textup{st}} a^\dagger])/\langle a^\dagger a\rangle^2$.
In the following we focus our analysis on the experimentally relevant regime $\Gamma_{\alpha\sigma}\ll\kappa$. This condition is consistent with previous reports on STM-induced light emission \cite{johansson1990, johansson1998} and ensures that the photon population in the plasmonic mode remains low, as photons are emitted much faster than they are generated by electron tunneling.
This allows us to truncate the Hilbert space to just the first few photonic states.
Labeling the system states as $|n, q\rangle$, where $n$ is the photon number and $q$ the electronic state, we restrict most analytical calculations to $n \leq 1$. 

The states up to the first-photon sector are shown in \Figref{fig:1}(b). Note that the state $\ket{1,e}$ belongs to the next photon sector, as it is only accessible from $\ket{1,0}$ and $\ket{1,d}$.
The equation of motion for the $7\times 7$ density matrix elements splits into \andres{31} separate blocks: 20 of dimension 1, 10 of dimension 2, and one of dimension 9.
This last one includes the 7 diagonal elements (populations) and the two off-diagonal terms between the states $\ket{0,e}$ and $\ket{1,g}$ (see \SM{S1.D}). 
These equations resemble the optical Bloch equations but include an effective driving term due to electron tunneling through the junction.
The relatively simple structure of the Liouvillian allows us to obtain transparent analytical expressions for the photon spectrum and the correlation function $g^{(2)}(\tau)$ discussed in the following.
In the numerics presented here we include all states containing up to three photons, and checked that the results are converged.
For the sake of simplicity in the following we present the results assuming $\Gamma_{\alpha\sigma} = \Gamma_\alpha$.

\textit{Electron and photon currents}---In \Figref{fig:1}(c,d) we show the electron and photon currents in the weak coupling and off-resonance regime as a function of the chemical potentials $(\mu_s-\varepsilon, \mu_t-\varepsilon)$ defined with respect to the ground state energy.
The structures in \Figref{fig:1}(c) reflect the population of states in the plasmon-molecule system. When current injection populates $\ket{0,e}$, photon emission occurs, shaping the features observed in \Figref{fig:1}(d).
We investigate the dependence of the currents on $\mu_s$ while fixing $\mu_t - \varepsilon = -0.5\omega_p$, as shown in \Figref{fig:1}(e,f).
It is observed that the first threshold for light emission occurs at an energy of $\mu_s - \varepsilon = \Delta$.
The width of the steps in \Figref{fig:1}(c--e) is set by $k_BT$. 
As shown in Ref.~\cite{schaeverbeke2019}, one needs to go beyond the Born-Markov expression for the current to obtain a width given by $\kappa$, as expected for $\kappa > k_B T$.
Here, we adopt the Born-Markov approximation, as our aim is not to resolve the step width, but to determine the energies at which light emission occurs.
We highlight a specific bias point, $(\mu_s, \mu_t) = \varepsilon + (1.4, -0.5) \omega_p$ (marked by $\star$ in \Figref{fig:1}).
This choice determines the transition rates $\Gamma_{\alpha, q' \rightarrow q}^\pm=\sum_{\sigma}\Gamma_{\alpha\sigma}f_\alpha^\pm(\epsilon_{qq'})|\bra{q}D^\pm_\sigma\ket{q'}|^2$, which in turn set the arrows observed in the many-body diagram in \Figref{fig:1}(b).
A detailed analysis for different coupling strengths and plateaus is provided in \SM{S3}.

We define the quantum yield as $\eta = I_{\rm ph}/I_{\alpha}$, which can be computed analytically at the $\star$ bias voltage point, yielding $\eta = \Gamma_{eg} / (2(\Gamma_t + \Gamma_{eg}))$, where $\Gamma_{eg}$ is given in \Eqref{eq:Gamma_eg}.
In \Figref{fig:2}(e), we show its dependence on $\Lambda/\kappa$ for both resonant and off-resonant cases. 
As shown in \Figref{fig:1}(b), saturation occurs when the states $\ket{0,e}$ and $\ket{1,g}$ are populated, allowing light emission at rate $\Gamma_{eg}$ or electron extraction at $\Gamma_t$. 
For $\Gamma_{eg} \gg \Gamma_t$, the yield reaches $\eta = 1/2$, due to equal partition of the electron tunneling between two channels, one leading to photon emission and the other not.
Conversely, when $\Gamma_{eg} \ll \Gamma_t$, electrons predominantly tunnel out elastically, yielding $\eta \approx \Gamma_{eg}/(2\Gamma_t) \propto \Lambda^2$.
The quadratic scaling arises because photon emission involves a second-order process: inelastic electron tunneling excites a plasmon, which rapidly decays by \andres{photo emission}.
The model predicts structures in the electronic and photonic currents, which are determined by the different tunneling thresholds.
These results set the stage for exploring the emitted light properties, which we address in the following.


\begin{figure} 
\centering
    \includegraphics[width=\columnwidth]{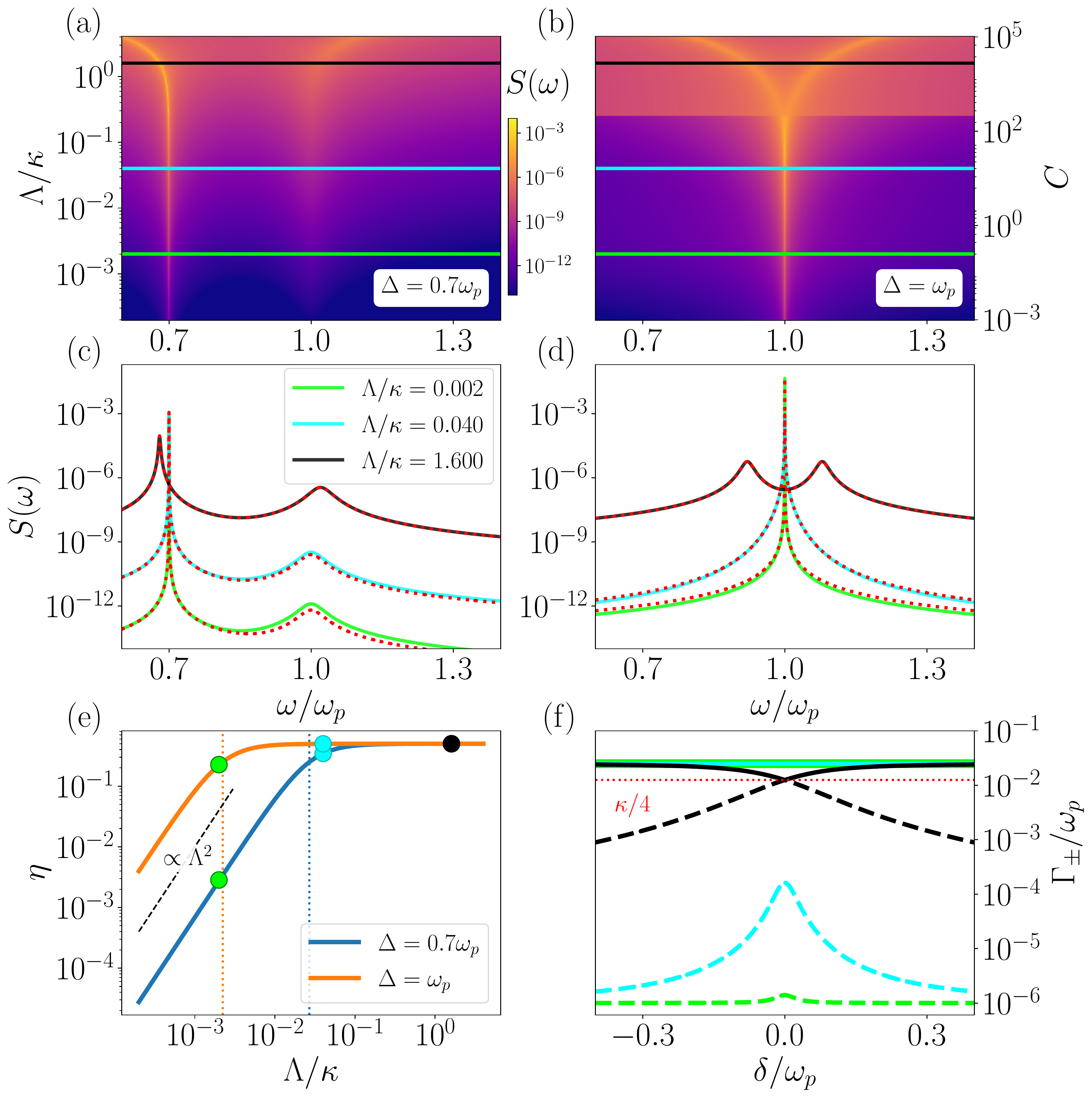}
    \caption{Computed emission spectra $S(\omega)$ for current-driven junctions.
    (a,b) Spectrum as a function of the emitted photon energy $\omega$ and plasmon-molecule coupling strength $\Lambda$ (or, equivalently, of the cooperativity $C$ on the right axis).
    (c,d) Slices of $S(\omega)$ corresponding to three distinct regimes characterized by $C$ and $\Lambda$, highlighting variations in resonance widths and positions.
    Dotted red lines correspond to the approximate, analytic expression \Eqref{eq:spectrum_star}.
    (e) Quantum efficiency $\eta$ of emitted photons as a function of $\Lambda$.
    The vertical dotted lines correspond to the condition $\Gamma_{eg}=\Gamma_t$, marking the transition between a $\eta\propto\Lambda^2$ scaling and the saturated regime $\eta=1/2$.
    (f) Linewidth $\Gamma_\pm$ of the two emission resonances as a function of the detuning $\delta=\Delta-\omega_p$ for the three coupling strengths (green, blue, black) considered in panels c-d.
    Dashed (full) lines represent the molecular (plasmonic) emission centered at $\Delta$ ($\omega_p$).
    The horizontal red dotted line corresponds to $\kappa/4$.
    The model parameters are the same as those of \Figref{fig:1} and the chemical potentials are fixed as indicated there by the $\star$.}
    \label{fig:2}
\end{figure}

\textit{Emission spectrum}---By projecting the expression for the emission spectrum on the basis of the (right and left) eigenvectors of the Liouvillian superoperator one obtains an expression as a sum of Lorentzians (see \SM{S1.F}).
Interestingly, in the limit of $\Gamma_{\alpha\sigma} \!\ll\! \kappa$, only three blocks contribute significantly to the spectrum. 
Two one-dimensional blocks, concerning $\braket{0,0|\rho|1,0}$ and $\braket{0,d|\rho|1,d}$, and one two-dimensional block for $\braket{0,g|\rho|1,g}$ and $\braket{0,g|\rho|0,e}$.
The reason can be understood as follows: photons are emitted from a state that is a linear combination of $\ket{0,e}$ and $\ket{1,g}$ to the state $\ket{0,g}$, or from state $\ket{1,0}$ ($\ket{1,d}$)  to the state $\ket{0,0}$ ($\ket{0, d}$).

With these simplifications, the spectrum at $\star$ bias is given by,
\begin{equation}
\begin{aligned}
\frac{(\Gamma_t + 2\Gamma_s)}{2\eta \Gamma_t \Gamma_s} S(\omega)\approx &~ \frac{\Gamma_t }{\kappa} L_{\omega_p,\kappa/2}(\omega)
    + \sum_{\mu=\pm}W_\mu L_{E_\mu, \Gamma_\mu}(\omega),
\label{eq:spectrum_star}
\end{aligned}
\end{equation}
where $L_{x, y}(\omega)=y/[(\omega-x)^2+y^2]/\pi$ is a Lorentzian function centered at $y$ with half-width at half max $y$, $W_\pm= \Lambda^2 {\rm Re}\big(1/[\Lambda^2-(\lambda_\pm-i \Delta+\Gamma_t)^2]\big)$ weight factors for the emission intensity, $\lambda_\pm =  i(\omega_p + \Delta)/2- \Gamma_t - \kappa/4 \pm \sqrt{(\kappa/2-i\delta)^2-4\Lambda^2}/2$ the leading eigenvalues of the Liouvillian, and $\delta = \omega_p -\Delta$ the detuning (see \SM{S2.C}).
The position and width are given by $E_\mu={\rm Re}(-i\lambda_\pm)$ and $\Gamma_\pm = {\rm Im}(-i\lambda_\pm)$, respectively.

In order to understand the evolution of the spectrum as a function of the different parameters we begin by expanding the Liouvillian eigenvalues $\lambda_\pm$ for $2\Lambda \ll |\kappa/2-i\delta|$. 
This leads to the following expressions: $\lambda_+=i\Delta-\Gamma_t-2\Lambda^2/(\kappa/2-i\delta)$ and
$\lambda_-=i\omega_p-\Gamma_t-\kappa/2 +2\Lambda^2/(\kappa/2-i\delta)$.
These correspond to the molecule and the plasmonic resonances, respectively.
The term proportional to $\Lambda^2$ gives a Lamb shift $2\delta\Lambda^2/(\kappa^2/4+\delta^2)$, pushing the resonances away from each other, and, more interestingly, a modification of their widths. 
The plasmonic resonance width $\Gamma_-$ is only weakly modified, since $\Gamma_t\ll\kappa$. 
On the contrary, the width of the molecular emission resonance reads
$\Gamma_+=\Gamma_t+\Gamma_{eg}/2$, where,
\begin{equation}
\Gamma_{eg}=C\frac{\kappa^2}{\kappa^2/4+ \delta^2} \widetilde\Gamma,    
\label{eq:Gamma_eg}
\end{equation}
represents the radiative decay rate from $|0,e\rangle$ to $|0,g\rangle$ (proportional to $\Lambda^2$) and 
$C = \Lambda^2 / \widetilde \Gamma\kappa$ is the cooperativity.
We define the electronic characteristic tunneling rate  $\widetilde{\Gamma}$ as the rate at which the states $\ket{0,e}$ and $\ket{1,g}$ perform a transition to either a state with 0 or 2 electrons (\SM{S1.F}).
Due to the specific bias configuration we are considering in \Figref{fig:1}(b), we have $\widetilde{\Gamma} = 2\Gamma_t$.
Note that \Eqref{eq:Gamma_eg} describes the emission of photons by the excited state of the TLS through the lossy plasmonic mode of the cavity, which can be viewed as a Purcell effect known in nanoplasmonics. However, the cooperativity $C$ does not coincide with the usual Purcell factor defined as the ratio between
the emission rate in the presence of the plasmonic environment and the emission rate in vacuum \cite{barreda2022}.
For $C>1$ the width of the molecular resonance is dominated by the coupling to the photons.
This is similar to what happens for instance in cavity optomechanics where for large $C$ the damping rate of the mechanical oscillator is dominated by the cavity back-action allowing cooling of the mechanical oscillator \cite{aspelmeyer2014}. 
We note that this condition does not require strong coupling, but only $\Lambda \gg \sqrt{2\Gamma_t\kappa}$, a much less stringent condition than $\Lambda\gg \kappa\gg\sqrt{2\Gamma_t\kappa}$. 
In this cooperativity regime, the first Lorentzian term in \Eqref{eq:spectrum_star} becomes negligible, as it is proportional to the population of the state $\ket{1,0}$ (see \Figref{fig:1}(b)).
The state $\ket{1,g}$ primarily decays via photon emission ($\Gamma_\pm$) or transitions to $\ket{1,0}$ ($\Gamma_t$), but since $\Gamma_\pm \gg \Gamma_t$ the population of the state  $\ket{1,0}$ is suppressed.  
The approximate expression in  \Eqref{eq:Gamma_eg} holds quite generally, but is not valid for $\delta\ll\kappa$ and $\Lambda>\kappa$.
In this case, the eigenvalues $\lambda_\pm$ reveal that for $\delta = 0$ and $\Lambda > \kappa/2$, both modes acquire identical damping rates of $\kappa/4$, while exhibiting a Rabi splitting of $2\Lambda$: $\lambda_\pm = i(\Delta + \omega_p)/2 \pm \Lambda - \kappa/4$. 

Both widths reduce to $\kappa/4$, which can be interpreted as two degenerate states, $\ket{0,e}$ and $\ket{1,g}$, that hybridize through coherent plasmon-molecule interactions and share a single emission channel of rate $\kappa/2$.
The width of the molecular mode thus increases with the coupling constant saturating at $\kappa/4$ in the strong coupling limit as we can see in \Figref{fig:2}(f), where we plot $\Gamma_\pm$.
Dashed lines correspond to the molecular resonance $\Gamma_+$ and solid lines to the plasmonic resonance $\Gamma_-$.
In the weak coupling the molecular resonance follows a Lorentzian shape in $\delta$, as given by \Eqref{eq:Gamma_eg}, with a minimum value at $2\Gamma_t$. 
The plasmonic resonance remains on the order of $\kappa$.
These results are illustrated in \Figref{fig:2}, where we plot the analytical expression from \Eqref{eq:spectrum_star} (dotted lines) and compare it with numerical results (solid lines).
The evolution of the spectrum $S(\omega)$ is shown as a function of the ratio $\Lambda/\kappa$ or $C$.
One can identify three regimes, (I) weak coupling and small cooperativity, (II) weak coupling but large cooperativity, with a parametrically broadened molecular peak, and (III) strong coupling ($C$ always large), with two symmetric Rabi peaks of width $\kappa/4$.
We note that a recent experimental study
variations of the emission line width has been observed for the same molecule in different geometric configurations  \cite{friedrich2024}.
%


\begin{figure} 
    \centering
    \includegraphics[width=\columnwidth]{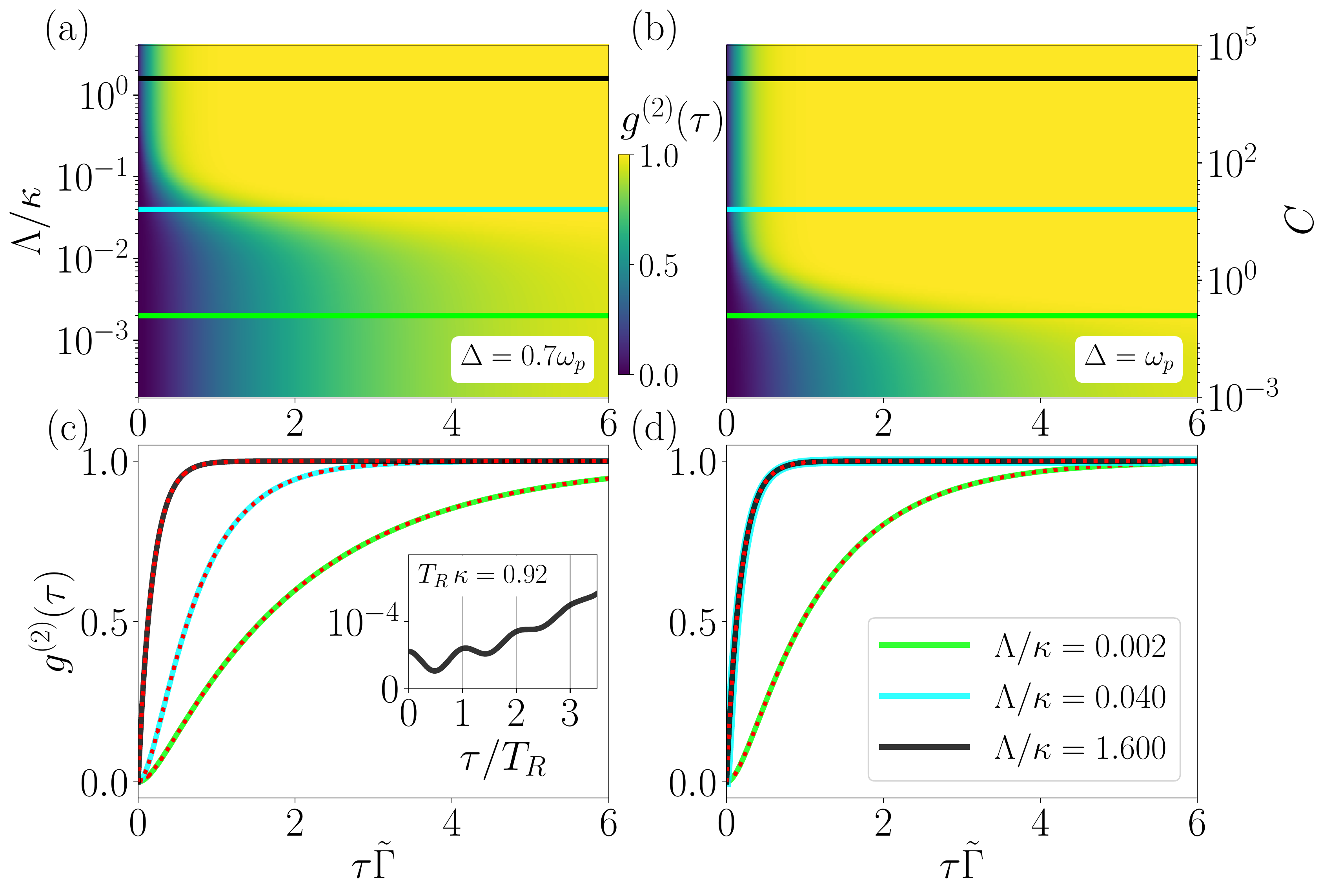}
    \caption{Analysis of the second-order photon correlation function $g^{(2)}(\tau)$.
    (a,b) On- and off-resonance cases, respectively, as a function of plasmon-molecule coupling strength $\Lambda$ and time delay $\tau$.
    (c,d) Slices of $g^{(2)}(\tau)$ corresponding to three distinct regimes characterized by $C$ and $\Lambda$, highlighting differences in antibunching properties and time evolution towards uncorrelated emission.
    Dotted red lines correspond to the approximate, analytic expression \Eqref{eq:g2_analytics}.
    The inset shows a zoom of $g^{(2)}(\tau)$ on very short timescales characterized by the Rabi period $T_R= 2\pi/\Omega_R=0.92\kappa$.
    The model parameters are the same as those of \Figref{fig:2}.}
    \label{fig:3}
\end{figure}

\textit{Photon correlations}---We begin with a simplified approach, expected to be valid for $\tau \gg \kappa^{-1}$.  
Since $\Gamma_t \ll \kappa $, we assume a time-scale separation between electrons and photons.  
On the slower electronic time scale, the plasmonic mode and electromagnetic environment can be treated as a structured bath with a spectral density:
$J(\omega) = (\Lambda^2/2\pi) \kappa/[(\omega-\omega_p)^2 + (\kappa/2)^2]$ \cite{garraway1997, petruccione2002}.
Using Fermi’s golden rule, the transition rate from the electronic excited state $\ket{e}$ to the ground state $\ket{g}$, with photon emission into the environment, is given by $\Gamma_{eg}=2\pi J(\Delta)$.
Note that this rate coincides with \Eqref{eq:Gamma_eg} found above from the spectrum calculation. 
This reduces the problem to a purely electronic Pauli master equation (see \SM{S4}), which can be solved analytically to obtain the following expression for the photon correlation function:
\begin{equation}
g^{(2)}(\tau) =
 1 + M e^{-(\Gamma_t + 2\Gamma_s)\tau}
- (1 + M )e^{-(\Gamma_t+\Gamma_{eg})\tau},
\label{eq:g2_analytics}
\end{equation}
where $M=(\Gamma_t + \Gamma_{eg})/(2\Gamma_s - \Gamma_{eg})$.
For any value of the rates entering this expression we find antibunching at short time, $g^{(2)}(0)=0$, and uncorrelated photons $g^{(2)}(\infty)=1$, at large time.
It is interesting to study the short time behavior.
There are naturally two time scales defined by the two exponentials. 
Surprisingly, for short time the behavior is always quadratic as the linear terms of the exponentials cancel, leaving $g^{(2)}(\tau)\approx(\Gamma_t + \Gamma_{eg})(2\Gamma_s+\Gamma_t)\tau^2/2$.
For simplicity, let us assume that $\Gamma_s \sim \Gamma_t$. 
The behavior of the function is then determined by the ratio $\Gamma_{eg}/\Gamma_t$.
For $\Gamma_t\gg \Gamma_{eg}$ (region I of the previous section) the quadratic region extends up to $\tau \sim 1/\Gamma_t$.
In contrast, for $\Gamma_{eg}\gg \Gamma_t$ (regions II and III) a linear behavior appears for $\tau>1/\Gamma_{eg}\ll 1/\Gamma_t$.

Numerical calculations of the full model agree with the predictions of this simple analytical approach.
Figure \ref{fig:3}(a,b) presents $g^{(2)}(\tau)$ as a function of the coupling strength for off- and on-resonance cases, respectively, while
\Figref{fig:3}(c,d) shows specific slices of $g^{(2)}(\tau)$ for each of the three coupling regions (I, II, III) introduced earlier.
The behavior is determined by the ratio $\Gamma_{eg} / \Gamma_t$, as predicted by the analytical theory.  
The inset in \Figref{fig:3}(c) displays the zoom to a short timescale on the order of $1/\kappa$ for strong coupling in the off-resonant configuration.
This regime is inaccessible to the analytical approach, but the full numerical model predicts oscillations at the Rabi frequency $\Omega_R = \sqrt{ \delta^2+4\Lambda^2}$.  
These features require a description of the plasmonic mode’s time evolution and cannot be captured by the structured environment approach.  
While these oscillations could provide an estimate of the coupling strength, their experimental observation is challenging due to the very short time scale $1/\kappa$.

\textit{Conclusions}---In summary, we have developed a general theoretical framework that captures both weak and strong coupling regimes, recovering results from structured bath models in the former case. Our analysis identifies three distinct regimes, each characterized by the cooperativity $C$, which manifest themselves in the excitation lifetimes observed in the emission spectrum $S(\omega)$ and the short-time behavior of photon correlations $g^{(2)}(\tau)$.

This framework provides a foundation for exploring more complex scenarios, with key ingredients that readily extend to broader settings.  Although our discussion has focused on scanning tunneling microscope (STM) current-driven light emission, the model is broadly applicable to other contexts involving current-induced photon emission, such as electron transport through molecular junctions or current driven plasmonic nanocavities.

\textit{Acknowledgments}---This research was supported by the Euskampus Transnational Common Laboratory \emph{QuantumChemPhys}, the IKUR Strategy under the collaboration agreement between the Ikerbasque Foundation and DIPC on behalf of the Department of Education of the Basque Government, the Spanish MCIN / AEI / 10.13039/501100011033 (PID2023-146694NB-I00, GRAFIQ), the Basque Department of Education (PIBA-2023-1-0021, TENINT), EUR Light S\&T Graduate Program (PIA3 Program ``Investment for the Future,'' ANR-17-EURE-0027), IdEx of the University of Bordeaux / Grand Research Program GPR LIGHT, and Quantum Matter Bordeaux.

\bibliography{bibliography}

\newpage
\clearpage

\begin{center}
    \Large \textbf{Supplementary Material}
\end{center}

\vspace{1em}

\noindent
\textbf{Abstract.} 
This supplementary material provides a detailed theoretical framework supporting our study on current-induced light emission in single-molecule tunnel junctions. We present the full microscopic Hamiltonian and derive the Lindblad master equation, including explicit density matrix elements and the Liouvillian structure. Key observables such as electronic and photonic currents, the photon emission spectrum, and second-order correlation functions are systematically derived. Additionally, we introduce an effective structured bath model, where we diagonalize the Hamiltonian and derive an effective master equation. These results offer deeper insights into the interplay between electronic transport and light emission.

\vspace{2em}

\maketitle

\tableofcontents

\section{Theoretical Framework}
\label{S.sec:qme}
In this appendix, we outline the theoretical framework and general formalism used in this work.

\subsection{Hamiltonian}
\label{S.sec:hamiltonian}

The open system under consideration is governed by the following Hamiltonian, which comprises three main terms: the central system $\op{H}_{\rm S}$, the environment (bath) $\op{H}_{\rm B}$, and the interaction between them, $\op{H}_{\rm SB}$.
We thus have:
\begin{equation}
    \op{H} = \op{H}_{\rm S} + \op{H}_{\rm B} + \op{H}_{\rm SB}.
    \label{S.eq:Hamiltonian}
\end{equation}
The Hamiltonian of the central system $\op{H}_{\rm S}$ (plasmon-molecule), introduced in the main text Eq.~(1), can be written as the sum of a non-interacting term, $\op{H}_0 = \op{H}_{\rm m}+\op{H}_{\rm p}$, and an interaction term, $\op{V}$.
They read
\begin{align}
\op{H}_{\rm p} =&~ \omega_p \opd{a} \op{a} ,
\label{S.eq:Hp}
\\
\op{H}_{\rm m} = &~  \varepsilon \opd{d}_g \op{d}_g+(\varepsilon +\Delta)  \opd{d}_e \op{d}_e+ U \opd{d}_g \opd{d}_e  \op{d}_e \op{d}_g
,
\label{S.eq:Hm}
\\
V=&~ \Lambda(\opd{a} \opd{d}_g \op{d}_e+a \opd{d}_e \op{d}_g).
\label{S.eq:V}
\end{align}
The description of each term has been given in the main text.
The non-interacting term ($\op{H}_0$) is diagonal, allowing us to express the eigenstates as $\ket{n,q}$, where $n$ is the number of photons and $q$ indicates the electronic states  $\{0,g,e,d\}$, as vacuum, ground state, excited state and doubly occupied state. 
The corresponding eigenenergies are given by $\epsilon_{nq} = n\omega_p+\epsilon_q$, with $\epsilon_q=0,\ \varepsilon, \ \varepsilon+\Delta$, and $2\varepsilon+\Delta+U$, respectively.

The system interacts with the electronic environment of the metallic leads and the electromagnetic environment of the propagating photons. 
This can be modeled by the 
following Hamiltonian for the bath: 
\begin{align}    
\op{H}_{\rm B} & = \op{H}_{\rm leads} + \op{H}_{\rm em} ,
\label{S.eq:HB}
\\
\op{H}_{\rm leads} & = \sum_{\alpha k} \epsilon_{\alpha k} \opd{c}_{\alpha k} \op{c}_{\alpha k} ,
\label{S.eq:Hleads}
\\
\op{H}_{\rm em} & = \sum_\nu \omega_\nu \opd{b}_\nu \op{b}_\nu,
\label{S.eq:Hem}
\end{align}
where  $\op{c}_{\alpha k}$ and  $\op{b}_\nu$ are the annihilation operators for the electrons in lead $\alpha=({\rm s},\ {\rm t}$) for substrate and tip, and the propagating photons, respectively.  
The electronic and photonic energy levels are given by $\epsilon_{\alpha k}$ and $\omega_\nu$. 

The interaction between system and bath is given by 
\begin{align}
\op{H}_{\rm SB} & =  \op{H}_{\rm tun} + \op{H}_{\rm p-em} ,
\label{S.eq:H_sb}
\\
 \op{H}_{\rm tun} & =  \sum_{\alpha k}\sum_{\sigma=g,e} \left[ t_{\alpha k\sigma} \opd{c}_{\alpha k} \op{d}_\sigma + t_{\alpha k \sigma}^* \opd{d}_\sigma \op{c}_{\alpha k} \right] , 
 \label{S.eq:Htun}
 \\
\op{H}_{\rm p-em}  &=  \sum_\nu\left[  g_\nu \opd{b}_\nu\op{a}  + g_\nu^* \opd{a}\op{b}_\nu  \right].
\label{S.eq:Hpem}
\end{align}
The first term represents the tunneling Hamiltonian between the reservoir and the two-level system, while the second term accounts for plasmonic losses. 
Specifically, $t_{\alpha k \sigma}$ denotes the electronic tunneling amplitudes, and $g_\nu$ describes the coupling strength between the plasmonic mode and the photonic reservoir.

\subsection{Lindblad master equation}
\label{S.sec:lindblad_master_equation}
The total density matrix of the system-bath, $\op{\rho}_T$, evolves according to the full Hamiltonian given by \Eqref{S.eq:Hamiltonian},
\begin{align}
 \dot{\op{\rho}}_T(t) & = -i[\op{H}, \op{\rho}_T(t)],
 \label{S.eq:rho_von_neumann}
\end{align}
where the density matrix describes the evolution  in the Hilbert space product of the system and environment Hilbert spaces.
We consider now the interaction picture with respect to $\op{H}'_0= \op{H}_{\rm S} + \op{H}_{\rm B}$ defining as usual
$
 \op{\rho}_T^I(t)  = 
 e^{i\op{H}'_0 t} \op{\rho}_T(t) e^{-i\op{H}'_0 t}$
and 
$\op{H}^I_{\rm SB}(t)=
e^{i\op{H}'_0 t} \op{H}^I_{\rm SB} e^{-i\op{H}'_0 t}
$ so that 
the \Eqref{S.eq:rho_von_neumann} becomes:
\begin{equation}
   \dot{\op{\rho}}_T^I(t) = -i[\op{H}^I_{\rm SB}(t),\op{\rho}_T^I(t)].
   \label{S.eq:rho_von_neumann_I}
\end{equation}
Integrating \Eqref{S.eq:rho_von_neumann_I}, we obtain $\dot{\op{\rho}}_T^I(t) = \op{\rho}_T^I(0)- i\int_0^t dt'[\op{H}^I_{\rm SB}(t'),\op{\rho}_T^I(t')]$. Substituting this expression for $\op{\rho}_T^I(t)$ back into the right-hand side of \Eqref{S.eq:rho_von_neumann_I} and tracing out the environmental degrees of freedom, we arrive at:
\begin{equation}
     \dot{\op{\rho}}^I(t) = -\int_0^t dt'\textup{Tr}_{\rm B}[\op{H}^I_{\rm SB}(t),[\op{H}^I_{\rm SB}(t'),\op{\rho}^I(t')]].
     \label{S.eq:rho_von_neumann_I_second_order}
\end{equation}
Where $\op{\rho}^I(t)$ is the reduced density matrix for the system and we have made use of 
the \textit{Born approximation}, by assuming %
$\op{\rho}_T(t) \approx \op{\rho}_{\rm leads} \otimes \op{\rho}_{\rm em} \otimes \op{\rho}(t)$.
Since the bath density matrix is in thermal equilibrium, $\textup{Tr}_{\rm B}[\op{H}^I_{\rm SB}(t),\op{\rho}_T^I(0)]=0$.
Next, we apply the \textit{Markov approximation}, which assumes that the environment correlation functions decay much faster than the typical timescale of evolution of the density matrix in the interaction picture.
One can then substitute $\rho(t')$ with $\rho(t)$,  replace the integral over $t' = t-\tau$ by integrating in $\tau$ and extent the integration to infinity.

This leads to
\begin{widetext}
\begin{equation}
\begin{aligned}
    \dot{\op{\rho}}^I(t) =& 
    \sum_{\sigma\sigma'}\int_0^\infty d\tau\Bigg\{\mathcal{C}_{\sigma\sigma'}^+(\tau)[\opd{d}_{\sigma', I}(t-\tau)\op{\rho}^I(t),\op{d}_{\sigma, I}(t)]   
    + \mathcal{C}_{\sigma'\sigma}^-(\tau)[\op{d}_{\sigma', I}(t-\tau)\op{\rho}^I(t),\opd{d}_{\sigma, I}(t)]\Bigg\}
    \\
   & +\int_0^\infty d\tau\Bigg\{ \mathcal{K}^+(\tau)[\opd{a}_I(t-\tau)\op{\rho}^I(t), \op{a}_I(t)] 
   +\mathcal{K}^-(\tau)[\op{a}_I(t-\tau)\op{\rho}^I(t), \opd{a}_I(t)]\Bigg\}+\textup{h.c}.
    \label{S.eq:me_int_picture}
\end{aligned}
\end{equation}
\end{widetext}
Here, we define the bath self-correlation functions,
\begin{align}
    \mathcal{C}^+_{\sigma\sigma' }(\tau) & = \sum_{\alpha k}t_{\alpha k\sigma}t^*_{\alpha k\sigma'}\langle c_{\alpha k}^\dagger(\tau)c_{\alpha k}(0) \rangle,
    \label{S.eq:c_plus_sigmasigma'}
    \\
      \mathcal{C}^-_{\sigma\sigma' }(\tau) & = \sum_{\alpha k}t_{\alpha k\sigma}^*t_{\alpha k\sigma'}\langle c_{\alpha k}(\tau)c_{\alpha k}^\dagger(0) \rangle,
        \label{S.eq:c_minus_sigmasigma'}
\\
    \mathcal{K}^+(\tau) & = \sum_{\nu}|g_\nu|^2\langle b_\nu^\dagger(\tau)  b_{\nu}\rangle,
    \label{S.eq:k_plus_sigmasigma'}
    \\
    \mathcal{K}^-(\tau) & = \sum_{\nu}|g_\nu| \langle b_\nu(\tau)  b_{\nu'}^\dagger\rangle.
    \label{S.eq:k_minus_sigmasigma'}
\end{align}
Since the environment's density matrix is in thermal equilibrium, the bath correlation function follows:
\begin{align}
    \langle c_{\alpha k}^\dagger(\tau)c_{\alpha k}(0) \rangle &= f_\alpha^+(\epsilon_{\alpha k})e^{ i\epsilon_{\alpha k}\tau},
    \label{S.eq:corr_plus}
    \\
    \langle c_{\alpha k}^\dagger(\tau)c_{\alpha k}(0) \rangle & =f_\alpha^-(\epsilon_{\alpha k})e^{ -i\epsilon_{\alpha k}\tau} ,
    \label{S.eq:corr_minus}
    \\
    \langle b_\nu^\dagger(\tau)  b_{\nu}\rangle & = n_B^+(\omega_\nu)e^{ i\omega_\nu\tau} ,
    \label{S.eq:korr_plus}
        \\
    \langle b_\nu(\tau)  b_{\nu}^\dagger\rangle & = n_B^-(\omega_\nu) e^{ -i\omega_\nu\tau} .
    \label{S.eq:korr_minus}
\end{align}
We have introduced the Fermi distribution functions for electron addition, $f_\alpha^+(\epsilon_{\alpha k}) = f_\alpha(\epsilon_{\alpha k})$, and hole addition, $f_\alpha^-(\epsilon_{\alpha k}) = 1 - f_\alpha(\epsilon_{\alpha k})$, where
$f_\alpha(\epsilon_{\alpha k}) = \left(e^{(\epsilon_{\alpha k} - \mu_\alpha)/k_B T} + 1\right)^{-1}$
is the Fermi-Dirac distribution in electrode $\alpha$, with $\mu_\alpha$ denoting the chemical potential, $T$ the temperature, and $k_B$ the Boltzmann constant. Similarly, for photons, we define $n_B^+(\omega_\nu) = n_B(\omega_\nu)$ and $n_B^-(\omega_\nu) = 1 + n_B(\omega_\nu)$, where
$n_B(\omega_\nu) = \left(e^{\omega_\nu/k_B T} - 1\right)^{-1}$
is the Bose-Einstein distribution.

Then we can insert Eqs.~\eqref{S.eq:corr_plus}, \eqref{S.eq:corr_minus}, \eqref{S.eq:korr_plus} and \eqref{S.eq:korr_minus} into Eqs.~\eqref{S.eq:c_plus_sigmasigma'},  \eqref{S.eq:c_minus_sigmasigma'}, \eqref{S.eq:k_plus_sigmasigma'} and \eqref{S.eq:k_minus_sigmasigma'} to obtain the self-correlations functions of the baths,
\begin{align}
    \mathcal{C}^\pm_{\sigma\sigma' }(\tau) & = \sum_{\alpha k}t_{\alpha k\sigma}t_{\alpha k\sigma'}^* e^{ \pm i\epsilon_{\alpha k}(\tau)}f_\alpha^\pm(\epsilon_{\alpha k}),
    \label{S.eq:lead_selfcorrelation}
\\
    \mathcal{K}^\pm(\tau) & = \sum_{\nu}|g_\nu|^2e^{\pm i\omega_\nu\tau} n_B^\pm(\omega_\nu).
    \label{S.eq:em_selfcorrelation}
\end{align}

In \Eqref{S.eq:me_int_picture}, the label $I$ refers to the interaction picture with respect to $H_{\rm S}$. 
Since the coupling between the system and the bath is weak, we can neglect the effect of the plasmon–molecule interaction $\Lambda$. However, when we return to the Schrödinger picture, we retain the interaction in the coherent evolution: $\dot{\op{\rho}}(t) = -i[\op{H}_{\rm S}, \op{\rho}(t)] + e^{-i\op{H}_{\rm 0}t} \dot{\op{\rho}}^I(t)e^{i\op{H}_{\rm 0}t}
$. 
This is known as the \textit{independent velocities approximation} (Sec. V.A.2 in \cite{cohen-tannoudji_atom_1998}). Under this assumption we can write the system operators in the interaction picture respect to $H_0$ as: 
\begin{align}
D_{I,\sigma}^\pm(t) & = \sum_{q_1q_2}e^{\pm i\epsilon_{q_1q_2}t}(D_{\sigma}^\pm)_{q_2}^{q_1},
\label{S.eq:D_I_H0}
\\
a_I(t) & = e^{-i\omega_pt}a,
\label{S.eq:a_I_H0}
\end{align}
where we introduced the operators,
\begin{equation}
    (D_\sigma^\pm)^{q_1}_{q_2}=\sum_n \ketbra{n,q_1} {n,q_1} D_\sigma^\pm \ketbra{n,q_2}{n,q_2},
\label{S.eq:D_pm}
\end{equation}
where $D_\sigma^- = \op{d}_\sigma$ and  $D_\sigma^+ = \op{d}^\dagger_\sigma$. 
Thus we rewrite \Eqref{S.eq:me_int_picture} in this form, yields the following expression:
\begin{widetext}
\begin{equation}
\begin{aligned}
   \dot{\op{\rho}}^I(t) 
    = &~\frac{1}{2}\sum_{q_1q_2q_3q_4}e^{i(\epsilon_{q_1q_2}-\epsilon_{q_4q_3})t}
\sum_{\alpha\sigma\sigma'}   \Gamma_{\alpha\sigma\sigma'}\sum_{\mu=\pm}f^\mu_\alpha(\epsilon_{q_1q_2})[(D^\mu_{\sigma'})_{q_2}^{q_1}\op{\rho}^I(t),(D^{\bar{\mu}}_{\sigma})_{q_4}^{q_3}]
   \\
   &+\frac{1}{2}\kappa n_B^+(\omega_p)[\opd{a}\op{\rho}^I(t), \op{a}]
 +\kappa n_B^-(\omega_p)[\op{a}\op{\rho}^I(t), \opd{a}]+\textup{h.c}.
\label{S.eq:me_int_picture_eigen}
\end{aligned}
\end{equation}
\end{widetext}
We have neglected the principal part in  the integral over $\tau$ as it involves a small shift in the coherent evolution of the density matrix, then: 
\begin{align}
 \int_0^\infty d\tau \mathcal{C}^\pm_{\sigma\sigma'}(\tau)e^{\mp i\omega\tau} & \approx  \frac{1}{2}\Gamma_{\alpha\sigma\sigma'}f^\pm_\alpha(\omega),
    \\
  \int_0^\infty d\tau \mathcal{K}^\pm(\tau)e^{\mp i\omega\tau}& \approx \frac{1}{2}\kappa n_B^+(\omega).
\end{align}
This gives the  tunneling rates that are defined as $\Gamma_{\alpha\sigma\sigma'} = 2\pi t_{\alpha\sigma}^*t_{\alpha\sigma'}\rho_\alpha$ where $\rho_\alpha$ the density of states in the lead $\alpha$. Similarly, the damping rate is $\kappa = 2\pi|g_\nu|^2\rho_{\rm em}$ with $\rho_{\rm em}$ denoting the electromagnetic environment's density of states.

The solution of \Eqref{S.eq:me_int_picture_eigen} may not always preserve the positivity and normalization of the density matrix during time evolution see for instance discussion in  Sec. 4.3 of \cite{schlosshauer2007}).
When we neglect the interaction $\Lambda$ the two systems, electrons and photons, are independent, and for each subsystem the typical evolution rate of the density matrix in the interaction picture is much smaller than the energy gap between any two states. 
This allows to use the
\textit{Secular approximation} (see Sec. 3.3.1 of \cite{petruccione2002}) to derive the disspators.
This approximation removes fast-oscillating terms in the master equation, which average to zero when the timescale of the density matrix evolution in the interaction picture is much longer than the oscillation timescales. 
Specifically, this condition is expressed as $\Gamma_{\alpha\sigma} \ll |\varepsilon_{q_1q_2} - \varepsilon_{q_4q_3}|$, where $\varepsilon_{q_1q_2} - \varepsilon_{q_4q_3}$ are the phase factors from \Eqref{S.eq:me_int_picture_eigen}.
We retain only the energy differences $\epsilon_{q_1}= \epsilon_{q_3}= \epsilon_{q}$ and $\epsilon_{q_2}=\epsilon_{q_4}=\epsilon_{q'}$, as other terms, even when $\sigma\neq\sigma'$, oscillate rapidly. 
Once this approximation is performed one can show that the 
dissipator are in Lindblad form, that ensure conservation of probability and positivity of the density matrix.
The general Lindblad form can be written as 
$\dot{\op{\rho}}(t) = -i[\op{H}_{\rm S}, \op{\rho}(t)] + \sum_{ij} k_{ij}[\opd{A}_j \op{\rho} \op{A}_i - \{\op{A}_i \opd{A}_j, \op{\rho}\}/2]$.

The photonic part of \Eqref{S.eq:me_int_picture_eigen} is already in Lindblad form, so no further modifications are needed for this part. 

After this approximation we finally transform \Eqref{S.eq:me_int_picture_eigen} to the Schrödinger picture.
%
%
The resulting equation of motion can be written as $\dot{\rho} = \mathcal{L}[\rho]$
where
\begin{equation} 
    \mathcal{L}  = \mathcal{L}_{\rm c}  + \mathcal{L}_{\rm e}  + \mathcal{L}_{\rm ph},
    \label{S.eq:full_liouvillian}
\end{equation}
with,
\begin{align}
      \mathcal{L}_{\rm c} \rho & = -i[\op{H}_{\rm S}, \op{\rho}],
      \label{S.eq:L_coherence}
    \\
    \mathcal{L}_{\rm e} & = \mathcal{L}_{\rm e}^+ + \mathcal{L}_{\rm e}^-,
    \label{S.eq:electronic_dissipator}
    \\
    \mathcal{L}_{\rm e}^\pm & = \sum_{\alpha qq'\sigma}   \Gamma_{\alpha\sigma}f^\pm_\alpha(\epsilon_{qq'})\mathcal{D}[(D_\sigma^\pm)_{q'}^{q}],
    \label{S.eq:electronic_dissipator_pm}
    \\
      \mathcal{L}_{\rm ph} & =  \kappa n_B^+(\omega_p)\mathcal{D}[\op{a}^\dagger] +\kappa n_B^-(\omega_p) \mathcal{D}[a],
    \label{S.eq:photonic_dissipator}
\end{align}
where $\mathcal{D}[X]\op{\rho} =\op{X}\op{\rho}\opd{X}- \{\opd{X}\op{X}, \op{\rho}\}/2$.

\subsection{Density matrix elements}
\label{S.sec:density_matrix}
We define the density matrix elements as $\rho_{n_1q_1,n_2q_2} = \bra{n_1,q_1} \rho \ket{n_2,q_2}$. 
By projecting the master equation onto this basis, $\dot{\rho}_{n_1 q_1,n_2q_2} = \bra{n_1, q_1} \mathcal{L} \op{\rho} \ket{n_2,q_2}$, we obtain the equations of motion for the matrix elements. 
We begin with the coherent evolution in \Eqref{S.eq:L_coherence}, where the system Hamiltonian is decomposed into three contributions according to the definition of $H_{\rm S}$.
Given that $\ket{n,q}$ are eigenstates of $\op{H}_0$, we write:
\begin{widetext}
\begin{equation}
\begin{aligned}
\bra{n_1, q_1} \mathcal{L}_{\rm c} \op{\rho} \ket{n_2,q_2}
 =-i\bra{n_1,q_1}  [\op{V}, \op{\rho}]\ket{n_2, q_2} 
     + i\epsilon_{q_2q_1}\rho_{n_1q_1,n_2q_2}
  + i\omega_p(n_2-n_1)\rho_{n_1q_1,n_2q_2}.
     \label{S.eq:L_coherente_matrix_elements}
\end{aligned}
\end{equation}

We express the total electronic dissipator in \Eqref{S.eq:electronic_dissipator} incorporating \Eqref{S.eq:D_pm},
\begin{equation}
    \mathcal{L}_{\rm e}\op{\rho} = \sum_{nqq'}\Gamma_{q'q}\left[P_{nq'}\ketbra{n,q}{n,q}-\frac{1}{2}\{\ketbra{n,q'}{n,q'}, \op{\rho}\}\right],
    \label{S.eq:electronic_dissipator_rates}
\end{equation}
where we define the total electron tunneling transition rates,
\begin{align}
\Gamma_{q'q} =&~ \sum_\alpha\Gamma_{\alpha, q'\rightarrow q}^+ + \Gamma_{\alpha, q'\rightarrow q}^-,
\\
 \Gamma_{\alpha, q' \rightarrow q}^\pm =&~ \sum_{\sigma}\Gamma_{\alpha\sigma}f_\alpha^\pm(\epsilon_{qq'})|\bra{q}D^\pm_\sigma\ket{q'}|^2.
\label{S.eq:tunneling_rate_plus}
\end{align}
We calculate now the matrix elements of \Eqref{S.eq:electronic_dissipator_rates}:
\begin{equation}
\begin{aligned}
\bra{n_1, q_1} \mathcal{L}_{\rm e} \op{\rho} \ket{n_2,q_2}
    & =  \sum_{nq'q}\delta_{n_2,n}\delta_{n_1,n}\delta_{q_1,q}\delta_{q_2,q}\Gamma_{q'q}P_{n,q'}
  -\sum_{nq'q}\frac{1}{2}\delta_{n_1, n}\delta_{q_1,q'}\Gamma_{q'q}\rho_{nq',n_2q_2}
 -\frac{1}{2}\sum_{nq'q}\delta_{n, n_2}\delta_{q',q_2}\Gamma_{q'q}\rho_{n_1q_1,nq'}
    \\
    &=\delta_{n_1, n_2}\delta_{q_2,q_1}\sum_{q}\Gamma_{qq_1}P_{n_1q}
   -\frac{1}{2}\sum_{q}[\Gamma_{q_1q} + \Gamma_{q_2q}]\rho_{n_1q_1,n_2q_2},
    \label{S.eq:electronic_dissipator_matrix_elements}
\end{aligned}
\end{equation}
where $\delta_{ij}$ is the Kronecker delta. 
The derivation of the matrix elements of \Eqref{S.eq:photonic_dissipator} is given for instance in Sec 6.2.1 of \cite{walls2008}:
\begin{equation}
\begin{aligned}
\bra{n_1, q_1} \mathcal{L}_{\rm ph} \op{\rho} \ket{n_2,q_2}=&~ \kappa n_B^+(\omega_p)\sqrt{n_1n_2}\op{\rho}_{(n_1-1)q_1,(n_2-1)q_2} 
  -\kappa n_B^+(\omega_p)\frac{(n_1+n_2+2)}{2}\op{\rho}_{n_1q_1,n_2q_2}
    \\
    & + \kappa n_B^-(\omega_p)\sqrt{(n_1+1)(n_2+1)}\op{\rho}_{(n_1+1)q_1,(n_2+1)q_2} 
   -\kappa n_B^-(\omega_p)\frac{(n_1+n_2)}{2}\op{\rho}_{n_1q_1,n_2q_2}.
    \label{S.eq:photonic_dissipator_matrix_elements}
\end{aligned}
\end{equation}
Since we are working at optical frequencies, where $n_B(\omega_p) =0$, the dissipator for the photonic in \Eqref{S.eq:photonic_dissipator}  part simplifies to $\mathcal{L}_{\rm ph}  = \kappa\mathcal{D}[a]$, which is the expression used in the main text.

We can finally combine Eqs.~\eqref{S.eq:L_coherente_matrix_elements}, \eqref{S.eq:electronic_dissipator_matrix_elements} and \eqref{S.eq:photonic_dissipator_matrix_elements} with $n_B(\omega_p) = 0$ to write the matrix elements of the density matrix:

\begin{equation} 
\begin{aligned} 
    \dot{\rho}_{n_1q_1,n_2q_2} & =\left[i\epsilon_{q_2q_1} -  \frac{1}{2}\sum_{q}(\Gamma_{q_1q} +\Gamma_{q_2q})\right]\rho_{n_1q_1,n_2q_2} +\delta_{n_1, n_2}\delta_{q_2,q_1}\sum_{q}\Gamma_{qq_1}P_{n_1,q}
    \\
    &~+\left[i(n_2-n_1)\omega_p-\frac{\kappa(n_1+n_2)}{2}\right]\rho_{n_1q_1,n_2q_2} +\kappa \sqrt{(n_1+1)(n_2+1)}\rho_{(n_1+1)q_1,(n_2+1)q_2}
    \\
    &~-i\bra{n_1,q_1}[V,\rho]\ket{n_2,q_2}.
    \label{S.eq:rho_matrix_elements}
\end{aligned}
\end{equation}
\end{widetext}


\subsection{Liouvillian structure}
\label{S.sec:liouvillian_structure}
In the limit $\Gamma_{\alpha\sigma}\ll \kappa$, it is reasonable to assume that states with more than one photon are rarely populated. The state $\ket{1,e}$ is also neglected, as it can only be populated from the states $\ket{1,0}$ and $\ket{1,d}$. However, these states decay very quickly with rate $\kappa$ to the states $\ket{0,0}$ and $\ket{0,d}$, respectively. Note that photons cannot be created from the bath, so the transition from $\ket{0,e}$ to $\ket{1,e}$  is not possible.
Under this assumptions the density matrix can be effectively reduced to the Hilbert space spanned by the 7 states $\ket{0,0}$, $\ket{1,0}$, $\ket{0,g}$, $\ket{0,e}$, 
$\ket{1,g}$, $\ket{0,d}$ and $\ket{1,d}$.
The 49 matrix elements of the density matrix can be regarded as a vector on which the super-operator $\cal L$ acts.
One can formally write the master equation as   $\underline{\dot{\rho}} = \check{\mathcal{L}}\underline{\rho}$,
where $\check{\mathcal{L}}$ is a matrix of dimension 49.
Since the dissipative part of the Liouville equation has a Lindblad form, the equations of motions have a relatively simple structure.
We begin by discussing the populations. 
One finds that they are only coupled among themselves, except when the last term in \Eqref{S.eq:rho_matrix_elements} is nonzero. 
This occurs for the states $\bra{0,e}[V,\rho]\ket{0,e} = \Lambda(\rho_{1g, 0e} - \rho_{0e,1g})$ and $\bra{1,g}[V,\rho]\ket{1,g} = \Lambda(\rho_{0e, 1g} - \rho_{1g,0})$. Additionally, writing the evolution equations for the coherences $\rho_{1g,0e}$ and $\rho_{0e,1g}$ using \Eqref{S.eq:rho_matrix_elements}, we find that they, in turn, depend on the populations $P_{0e}$ and $P_{1g}$.
Consequently, if we define $\underline{\rho}_9 = (P_{00}, P_{10}, P_{0g}, P_{0e}, P_{1g}, P_{0d}, P_{1d},\rho_{1g,0e}, \rho_{0e,1g})^T$, we identify a $9\times 9$ block within the Liouvillian such that the evolution of this subspace is governed by $\dot{\underline{\rho}}_9= \check{\mathcal{L}}_{9\times9}\underline{\rho}_9$. Let us define $\Gamma_{q\rightarrow} = \sum_{q'}\Gamma_{qq'}$,
\begin{widetext}
\begin{equation}
\check{\mathcal{L}}_{9\times 9}=
\begin{pmatrix}
-\Gamma_{0\rightarrow} &   \kappa& \Gamma_{g0} & \Gamma_{e0} &  0& 0 &0 & 0 &0
\\
0& - \kappa &  0 & 0 & \Gamma_{g0} & 0 & 0& 0 &0
\\
\Gamma_{0g} &  0&-\Gamma_{g\rightarrow }&0 & \kappa & \Gamma_{dg} & 0 & 0&0
\\
\Gamma_{0e} & 0 & 0 &-\Gamma_{e\rightarrow} & 0 & \Gamma_{de} & 0 & -i\Lambda & i\Lambda
\\
0 & \Gamma_{0g} &  0& 0 & - \kappa & 0 &  \Gamma_{dg} &  i\Lambda & -i\Lambda
\\
0 & 0 & \Gamma_{gd} &  \Gamma_{ed} &  0 & -\Gamma_{d\rightarrow} & 0 & 0 & 0
\\

0  & 0 & 0 & 0 & \Gamma_{gd} &  0  & -\kappa  & 0& 0
\\
0 & 0 & 0& -i\Lambda & i\Lambda & 0 & 0 & -i\delta - \kappa/2 & 0
\\
0 & 0 & 0& i\Lambda & -i\Lambda & 0 & 0 & 0&i\delta - \kappa/2 
\label{S.eq:block9x9}
\end{pmatrix} ,
\end{equation}
\end{widetext}
where $\delta=\omega_p-\Delta$ the detuning.
In this equation, we have made the approximation $\Gamma_{qq'} + \kappa \approx \kappa$.

It is interesting to note that the equations above that involve $\ket{0,e}$ and $\ket{1,g}$ closely resemble the optical Bloch equations (See Sec $\rm V.A.3$ of \cite{cohen-tannoudji_atom_1998}).

Next, we analyze the coherences. 
When writing the coherence terms, we can neglect the contribution $\kappa \sqrt{(n_1+1)(n_2+1)} \rho_{(n_1+1)q_1, (n_2+1)q_2}$ in Eq.~\eqref{S.eq:rho_matrix_elements}. Using the Cauchy-Schwarz inequality, we have $|\braket{x|y}|^2\leq\braket{x|x}\braket{y|y}$. We choose $\ket{x}=\rho^{1/2}\ket{i}$ and $\ket{y}=\rho^{1/2}\ket{j}$ thus, $|\rho_{ij}|^2 < \rho_{ii} \rho_{jj}$. In the single-photon sector, this applies to terms such as $|\rho_{1q,1q'}|^2 < \rho_{1q,1q} \rho_{1q',1q'}$ with $q = 0, g, d$ and $q \neq q'$. All these terms involve at least the states $\ket{1,0}$ and $\ket{1,d}$, which remains small since these states decay rapidly at a rate $\kappa$. 
Under this considerations the evolution \Eqref{S.eq:rho_matrix_elements} reveals that the interaction leads to the formation of $2 \times 2$ blocks for all matrix elements of the form $\underline{\rho}_{2\times 2}(n,q) = (\rho_{nq, 0e},  \rho_{nq, 1g})^T$.
Specifically, the Liouvillian contains five such $2 \times 2$ blocks, along with their complex conjugates, resulting in a total of ten.
The Liouvillian of all $2\times 2$ blocks such that $\dot{\underline{\rho}}_{2\times 2}(n,q) = \mathcal{L}_{2\times 2}(n,q)\underline{\rho}_{2\times 2}(n,q)$ is given by:
\begin{widetext}
\begin{equation}
\check{\mathcal{L}}_{2\times 2}(n,q)=
\begin{pmatrix}
i(\epsilon_{eq} + n\omega_p)-(\Gamma_{q\rightarrow}+\Gamma_{e\rightarrow}  +n\kappa)/2 & i\Lambda 
\\
i\Lambda & i[\epsilon_{gq} + (1-n)\omega_p]-(\Gamma_{q\rightarrow}+\Gamma_{g\rightarrow} +(1+n)\kappa)/2
\label{S.eq:blocks2x2}
\end{pmatrix} .
\end{equation}
\end{widetext}
For completeness here we provide the remaining $1 \times 1$ blocks:
\begin{align}
    \dot{\rho}_{00,10} & = [i\omega_p - \kappa/2] \rho_{00,10},
    \label{S.eq:L10}
\\
    \dot{\rho}_{00,0g} & = [i\epsilon_{g0} - (\Gamma_{0\rightarrow} + \Gamma_{g\rightarrow})/2] \rho_{00,0g},
\\
    \dot{\rho}_{00,0d} & = [i\epsilon_{d0} -(\Gamma_{0\rightarrow} + \Gamma_{d\rightarrow})/2] \rho_{00,0d},
\\
    \dot{\rho}_{00,1d} & = [i(\epsilon_{d0} + \omega_p) -\kappa/2] \rho_{00,1d},
\\
   \dot{\rho}_{10,0g} & = [i(\epsilon_{g0} -\omega_p) -\kappa/2] \rho_{10,0g},
\\
    \dot{\rho}_{10,0d} & = [i(\epsilon_{d0}-\omega_p) -\kappa/2] \rho_{10,0d},
\\
  \dot{\rho}_{10,1d} & = [i\epsilon_{d0} -\kappa] \rho_{10,1d},
\\
    \dot{\rho}_{0g,0d} & = [i\epsilon_{dg} -(\Gamma_{g\rightarrow} + \Gamma_{d\rightarrow})/2] \rho_{0g,0d},
\\
    \dot{\rho}_{0g,1d} & = [i(\epsilon_{dg}+\omega_p) -\kappa/2] \rho_{0g,1d},
    \\
    \dot{\rho}_{0d,1d} & = [i\omega_p - \kappa/2] \rho_{0g,1d},
     \label{S.eq:L1d}
\end{align}
Once again, we have neglected the tunneling rates, where we impose terms like \( \kappa + \Gamma_{qq'} \approx \kappa \).

It is worth noting that when calculating the steady-state of the density matrix, all coherences go to zero except for those appearing in the \Eqref{S.eq:block9x9}.

In summary, the Liouvillian decomposes into a single $ 9 \times 9$ block, 10 $2 \times 2$ blocks, and 20 $1 \times 1$ blocks. 
In these last two groups we included in the enumeration the complex conjugate equations.

\subsection{Currents}
\label{S.sec:currents}
We now aim to compute the average electronic and photonic currents.
Following Ref. \cite{marcos2010} this can be achieved by 
introducing counting fields for the electrons crossing the junctions and the photons emitted. 
In practice one can show that it is possible to define a modified Liouville operator such the equation of 
motion for the density matrix becomes:
\begin{equation}
    \dot\rho(\xi_\alpha,\chi) =   \mathcal{L}(\xi_\alpha, \chi) \rho(\xi_\alpha,\chi) ,
\end{equation}
where $\xi_\alpha$ and $\chi$ are the conting fields for the $\alpha$-lead and the photons, respectively. 
The number of, say, electrons that crossed from reservoir $\alpha$ is then just 
\begin{equation}
\langle N_\alpha \rangle 
=
\left. \partial_{ \xi_\alpha} \rho \right|_{\chi=0,\xi_s=0,\xi_t=0} \, .
\end{equation}
Similarly higher moments can be calculated.
Following \cite{emary2012} the counting field is introduced in the dissipator as follows:
\begin{equation}
    \mathcal{D}(\xi)[A]\op{\rho} =\op{A}\op{\rho}\opd{A}e^{i\xi} - \{\opd{A}\op{A}, \op{\rho}\}/2.
    \label{S.eq:liouvillian_xi_chi}
\end{equation}
From this definition it follows the definition of the the jump (super-)operator:
\begin{equation}
    \mathcal{J}   =  -i\left[\frac{\partial}{\partial \xi }\mathcal{D}(\xi)\right]_{\xi = 0}.
    \label{S.eq:jump_def}
\end{equation}
In our case we have 
\begin{equation}
    \mathcal{L}(\xi_\alpha, \chi)  =  \mathcal{L}_{\rm coh}+ \mathcal{L}^+_{\rm e}(\xi_\alpha) + \mathcal{L}^-_{\rm e}(-\xi_\alpha)  + \mathcal{L}_{\rm ph}(\chi),
\end{equation}
where the counting fields have been introduced in each term following \Eqref{S.eq:liouvillian_xi_chi}.

In this approach, the average currents are given by the expectation value of the jump operator \cite{marcos2010},
\begin{equation}
    I   =  \langle \mathcal{J} \rangle \,.
\end{equation}
Following this definition in \Eqref{S.eq:jump_def}, we obtain the jump operators for the electronic and photonic part from \Eqref{S.eq:liouvillian_xi_chi}:
\begin{align} 
{\cal J}_{e,\alpha}^\pm \rho 
& =
\sum_{\sigma qq'}\Gamma_{\alpha \sigma}
f_\alpha^\pm(\epsilon_{qq'}) 
(D_\sigma^\pm)^{q'}_{q}\rho (D_\sigma^\mp)^{q}_{q'},
\label{S.eq:electronic_jump}
\\
{\cal J}_{\rm ph} \rho&= \kappa a\rho a^\dagger.
\label{S.eq:photonic_jump}
\end{align}
Let us start with the electronic current, one we insert the expression \eqref{S.eq:D_pm} in \eqref{S.eq:electronic_jump},
\begin{equation}
    {\cal J}_{e,\alpha}^\pm \rho 
=
\sum_{nqq'}\Gamma_{\alpha, q'\rightarrow q}^\pm P_{nq'}\ketbra{n,q}{n,q},
\label{S.eq:electronic_jump_basis}
\end{equation}
where transition rate $\Gamma_{\alpha, q' \rightarrow q}^\pm$ was given in \Eqref{S.eq:tunneling_rate_plus}.
Using the definition of $I_\alpha= d\langle N_\alpha\rangle/dt$ one finds that the 
electronic current is given by the outgoing flux from lead $\alpha$ minus the incoming flux $I_{\alpha} = \tr{[\mathcal{J}^+_{e, \alpha} -\mathcal{J}^-_{e, \alpha} ]\op{\rho}^{\text{st}} }$.
This gives 
\begin{equation}
    I_\alpha = \sum_{nqq'} (\Gamma_{\alpha, q'\rightarrow q}^+ - \Gamma_{\alpha, q'\rightarrow q}^-)P_{nq'}.
    \label{S.eq:electronic_current}
\end{equation}
Similarly the photonic current reads
\begin{equation}
    I_{\rm ph} = \tr{{\cal J}_{\rm ph} \rho} = \kappa \langle n\rangle. 
\label{S.eq:photonic_current}
\end{equation}


\subsection{Photon emission spectrum}
\label{S.sec:photoemission_spectrum}

In analogy to the standard definition of emission spectra \cite{cohen-tannoudji_atom_1998}, we express the spectrum in terms of the two-time correlation function of the field operators. 
\begin{equation}
    S(\omega) = \frac{\kappa}{2\pi} \int_{-\infty}^{+\infty} d\tau  e^{-i\omega\tau} \langle \op{a}^\dagger(\tau) \op{a}(0) \rangle.
    \label{S.eq:spectrum_plasmonic}
\end{equation}

This equation shows that the emission spectrum is directly determined by the two-time correlation function of the plasmonic mode. The prefactor $\kappa/2\pi$ arises from the decay rate of the plasmonic excitation into the electromagnetic continuum, effectively determining the linewidth of the emission.
\label{S.sec:emission_spectrum}

The time correlation function can be computed using the quantum regression theorem \cite{cohen-tannoudji_atom_1998}, 
\begin{equation}
S(\tau>0) = \langle \opd{a}(\tau) a(0)\rangle = \text{Tr}\left(\opd{a} e^{\mathcal{L}\tau}a\op{\rho}^{\text{st}} \right).
\label{S.eq:first_order_corr_a_qrt}
\end{equation}
The density matrix $\op{\rho}$ of an $N$-dimensional Hilbert space is represented as an $N \times N$ matrix. By vectorizing it into a $1 \times N^2$ vector $\underline{\rho}$, the master equation can be rewritten as $\dot{\underline{\rho}} = \check{\mathcal{L}}\underline{\rho}$, where $\check{\mathcal{L}}$ is the Liouvillian superoperator of dimension $N^2 \times N^2$.

Since $\check{\mathcal{L}}$ is generally non-Hermitian ($\check{\mathcal{L}}\neq\check{\mathcal{L}}^\dagger$), it possesses distinct left and right eigenvectors, satisfying $\check{\mathcal{L}} v_j = \lambda_j v_j$ and $w_j^t\check{\mathcal{L}} = \lambda_j w_j^t$, respectively, with the orthogonality condition $(w_i, v_j) = \delta_{ij}$. The inner product is defined as $(w_i, v_j)=\sum_k (w_i)^*_k (v_j)_k$, and the completeness relation reads $\sum_j |v_j)(w_j| = \mathbb{1}$.

The conservation of the probability leads to the existence of a stationary solution with vanishing eigenvalue: $\lambda_0 = 0$.
The corresponding right eigenvector $v_0 = \underline{\rho}^{\text{st}} $ is stationary solution of the equation of motion.
The left eigenvector associated with this eigenvalue satisfies $(w_0, v_j) = \text{Tr}(v_j)$. 
This allows us to express the correlation function in \Eqref{S.eq:first_order_corr_a_qrt} as:
\begin{equation}
\begin{aligned}
    S(\tau) =&~ (w_0, \check{a}^\dagger e^{\mathcal{\check{L}\tau}}\check{a}\underline{\rho}^{\text{st}} )
    \\
    =&~ \sum_j e^{\lambda_j\tau}(w_0, \check{a}^\dagger v_j)(w_j,\check{a}\underline{\rho}^{\text{st}} ).
\end{aligned}
\end{equation}
As demonstrated in \Secref{S.sec:liouvillian_structure}, the Liouvillian $\check{\mathcal{L}}$ can be expressed in block form. This allows us to define $\mathcal{S}_M$ as the subspace of the operator space that remains invariant under the action of $\check{\mathcal{L}}$, i.e., $\check{\mathcal{L}}\mathcal{S}_M \subseteq \mathcal{S}_M$. Within each subspace, we introduce a local operator, enabling us to decompose the Liouvillian as $
\check{\mathcal{L}} = \check{\mathcal{L}}_0 \oplus  \check{\mathcal{L}}_1\oplus \dots\oplus \check{\mathcal{L}}_M$. This structure allows us to label the relevant elements in the correlation function using $M$, where $M$ corresponds to a specific block within the Liouvillian,
\begin{equation}
S(\tau) =\sum_{j M} e^{\lambda_j^M\tau}(w_0, \check{a}^\dagger v_j^M)(w_j^M,\check{a}\underline{\rho}^{\text{st}} ).
\label{eq:first-order-corr-a-liuv_space_blocks}
\end{equation}
Next we define the projections on the basis of the Hilbert space of the plasmonic operator as $a_{\alpha\beta}=\bra{\alpha}\op{a}\ket{\beta}$ and the same for the eigenvectors operators $v_i$  $(v_j)_{\alpha\beta}=\bra{\alpha}\op{v}_j\ket{\beta}$.
Taking into account the completeness relation we get 
$(av)_{\alpha\beta} =\sum_{\delta}  a_{\alpha \delta}v_{\delta \beta}$.
We now can write the correlation function as:
\begin{equation}
S(\tau) =\sum_{jM}e^{\lambda_j^M\tau}A_j^MB_j^M ,
\label{S.eq:first_corre_final}
\end{equation}
where 
\begin{align}
    A_j^M= &\sum_{\alpha\beta}(a^\dagger)_{\alpha\beta}(v_j^M)_{\beta\alpha},
\label{S.eq:first_corre_final_A}
\\
B_j^M=& \sum_{\gamma\delta\theta}(w_j^M)^*_{\gamma\delta}(a)_{\gamma\theta}(\rho^{\text{st}} )_{\theta\delta}.
\label{S.eq:first_corre_final_B}
\end{align}
We can now use this formalism to perform the calculation of the spectrum.

By integrating \Eqref{S.eq:spectrum_plasmonic} one obtains for the emission spectrum,
\begin{align}
S(\omega) =\frac{\kappa}{\pi}\textup{Re}\sum_{jM\neq 0}A_j^MB_j^M \frac{\Gamma_i^M}{(\omega-E_j^M)^2 +(\Gamma_j^M)^2}.
\label{S.eq:spectrum_lorentzians}
\end{align}
Regardless of the specific values of the coefficients $A_j^M$ and $B_j^M$, the resulting emission spectrum is given by a sum of Lorentzians, with each component weighted by the product of these coefficients. 
In \Eqref{S.eq:spectrum_lorentzians}
we have introduced the lorentzian half-width at half-maximum (HWHM) and resonance position:
\begin{align}
    \Gamma_j^M & = {\rm Im}(-i\lambda_j^M),
    \label{S.eq:hwhm}
\\
     E_j^M & = {\rm Re}(-i\lambda_j^M).
    \label{S.eq:peak_position}
 \end{align}

One can show that the contribution of the zero eigenvalue can be discarded, since there is no coherent contribution to the emitted light.
To see this, let us consider the most general case where the Liouvillian cannot be written in block form. Then, by taking into account \Eqref{S.eq:first_corre_final_A} and \Eqref{S.eq:first_corre_final_B}, we find that:
\begin{align}
A_0 =& \sum_{\alpha\beta}(a^\dagger)_{\alpha\beta} (\rho^{\text{st}} )_{\beta\alpha},
\\
 B_0 =& \sum_{\gamma\delta\theta}
 {(w_0)}_{\gamma\delta}(a)_{\gamma\theta} (\rho^{\text{st}} )_{\theta\delta}  .  
\end{align}
The second coefficient $B_0$ contains a Kronecker delta 
$(w_0)_{\gamma\delta} = \delta_{\gamma\delta}$ then $B_0 = \sum_{\gamma\theta}(a)_{\gamma\theta} (\rho^{\text{st}} )_{\theta\gamma}$
In the bare basis $\{n,q\}$:
\begin{align}
A_0 =& \sum_{\substack{nq\\n'q'}}\bra{n',q'}\opd{a}\ket{n,q}\bra{n,q}\op{\rho}^{\text{st}} \ket{n',q'},
\\
B_0=&\sum_{\substack{nq\\n'q'}}\bra{n',q'}\op{a}\ket{n,q}\bra{n,q'}\op{\rho}^{\text{st}} \ket{n',q'}.
\end{align}
As $\opd{a}\ket{n}=\sqrt{n+1}\ket{n+1}$ and $\op{a}\ket{n}=\sqrt{n}\ket{n-1}$ we obtain:
\begin{align}
A_0 =& \sum_{nq}\sqrt{n+1}\rho^{\text{st}} _{nq, (n+1)q},
\\
B_0=&\sum_{nq}\sqrt{n}\rho^{\text{st}} _{nq,(n-1)q}.
\end{align}
This implies that the zero eigenvalue will contribute if coherences with different photon numbers are present in the steady state. In the most general case, these coherences do not necessarily vanish. However, the only coherences that remains in the stationary regime are $\bra{0,e}\op{\rho}\ket{1,g}$ and $\bra{1,g}\op{\rho}\ket{0,e}$ (See \Secref{S.sec:liouvillian_structure}). Consequently, we can conclude that the contribution of this term will be zero in our further analysis.

Let us focus on the term $A_j^M$. The eigenvector $v_j^M$ can be expressed as $v_j^M = \sum_{nqn'q'\in M}(v_j^M)_{nq,n'q'}\ket{n,q}\bra{n',q'}$ where  $\ket{n,q}\bra{n',q'}$ forms the basis in which each block $M$ of the Liouvillian is written. These results indicate that only the blocks containing nonzero contributions in the summation of $A_j^M$ in \Eqref{S.eq:first_corre_final_A} will influence the final outcome.
The expression for $A_j^M$ reads:
\begin{equation}
A_j^M = \sum_{nq\in M }(v_j^M)_{nq,(n+1)q}\sqrt{n+1}.
\label{S.eq:AiM_matrix_elemtns}
\end{equation}
In the limit $\Gamma_{\alpha, \sigma } \ll \kappa$, the only nonzero terms in \Eqref{S.eq:AiM_matrix_elemtns} are those involving the matrix elements of the operators $\ket{0,0}\bra{1,0}$, $\ket{0,d}\bra{1,d}$ and $\ket{0,g}\bra{1,g}$. 

As demonstrated in \Secref{S.sec:liouvillian_structure}, the former corresponds to a $1 \times 1$ blocks, while the latter forms a $2 \times 2$ block within the Liouvillian.

The  $1\times 1$ blocks for $\rho_{00,10}$ and $\rho_{0d,1d}$ are given in Eqs.~\eqref{S.eq:L10} and \eqref{S.eq:L1d}. Since these blocks are already diagonal, the eigenvalues are directly obtained as $\lambda^{10} = \lambda^{1d} = i\omega_p - \kappa/2$. The HWHM and resonance frequency follow from Eqs.~\eqref{S.eq:hwhm} and \eqref{S.eq:peak_position}, yielding $\Gamma^{10} = \Gamma^{1d} = \kappa/2$ and $E^{1d} = E^{10} = \omega_p$. 

Next, we determine the coefficients. The trivial case $A^{10} = A^{1d} = 1$ follows from \Eqref{S.eq:AiM_matrix_elemtns}. Finally, from \Eqref{S.eq:first_corre_final_B}, we obtain:
\begin{align}
    B^{10} = &~ \sum_{\gamma\delta\theta}\braket{\gamma|0,0}\braket{1,0|\delta}a_{\gamma\theta}(\rho^{\text{st}} )_{\theta\delta} = P_{10}^{\text{st}}  ,
    \\
     B^{1d} = &~ \sum_{\gamma\delta\theta}\braket{\gamma|0,d}\braket{1,d|\delta}a_{\gamma\theta}(\rho^{\text{st}} )_{\theta\delta} = P_{1d}^{\text{st}} .
\end{align}

Inserting the definition of  $A^{10}, A^{1d}, B^{10}, B^{1d}$ , $\lambda^{0d}$ and $\lambda^{1d}$ in \eqref{S.eq:spectrum_lorentzians} one obtains:
\begin{align}
S_{1}(\omega) = &   \frac{\kappa}{\pi} \sum_{q=0,d} P_{1q}^{\text{st}}\frac{\kappa/2}{[\omega-\omega_p]^2+[\kappa/2]^2} .
\label{S.eq:peak_omegap}
\end{align}
We turn now to the two-dimensional block:  
$\underline{\rho}_{2\times 2}(0, g) = (\rho_{0g,0e}, \rho_{0g,1g})^T$.
The Liouvillian operator in this block reads from \Eqref{S.eq:blocks2x2}:
\begin{equation}
\check{\mathcal{L}}_{2\times 2}(0,g)=
\begin{pmatrix}

 i\Delta-\widetilde{\Gamma}/2 & i\Lambda 
\\
i\Lambda & i\omega_p- \Gamma_{g\rightarrow} -\kappa/2 
\label{S.eq:L2}
\end{pmatrix}.
\end{equation}
We have introduce the charge loss rate $\widetilde{\Gamma} = \Gamma_{g\rightarrow}+\Gamma_{e\rightarrow}$, that represents all possible ways of leaving both the ground and excited states—in other words, the departure from the first charge sector.
The eigenvalues read:
\begin{equation}
\begin{aligned}
\lambda_\pm =~& i\frac{(\omega_p+\Delta)}{2} - \frac{\widetilde{\Gamma}}{4} - \frac{\Gamma_{g\rightarrow}}{2} - \frac{\kappa}{4} 
\\
&\pm\frac{1}{2}\sqrt{[(\kappa +  \Gamma_{g\rightarrow} -\Gamma_{e\rightarrow})/2-i\delta]^2-4\Lambda^2}.
\label{S.eq:lambda.pm}
\end{aligned}
\end{equation}
In the regime where  $2\Lambda \ll |(\kappa +  \Gamma_{g\rightarrow} -\Gamma_{e\rightarrow})/2-i\delta|$, the eigenvalues simplify to 
\begin{align}
\lambda_+ & =  i\Delta-\frac{\widetilde{\Gamma}}{2}-\frac{\Lambda^2}{(\kappa +  \Gamma_{g\rightarrow} -\Gamma_{e\rightarrow})/2-i\delta},
\\
\lambda_-& = i\omega_p-\frac{\kappa}{2} -\Gamma_{g\rightarrow}+\frac{\Lambda^2}{(\kappa +  \Gamma_{g\rightarrow} -\Gamma_{e\rightarrow})/2-i\delta}.
\end{align}  
In the next step we assume that $(\kappa +  \Gamma_{g\rightarrow} -\Gamma_{e\rightarrow})/2\approx \kappa/2$.
The position of each resonance is given by \Eqref{S.eq:peak_position},
\begin{align}
    E_+ & = \Delta - \frac{\Lambda^2\delta}{(\kappa/2)^2+\delta^2},
    \\
      E_- &= \omega_p + \frac{\Lambda^2\delta}{(\kappa/2)^2+\delta^2}.
\end{align}
The second term in each resonance, proportional to $\Lambda^2$, represents the repulsion between the molecule and the plasmonic. This contribution is known as the Lamb shift.
The corresponding widths are obtained from \Eqref{S.eq:hwhm}:
\begin{align}
    \Gamma_+ &\approx \frac{1}{2}(\widetilde\Gamma + \Gamma_{eg}),
    \\
    \Gamma_- & \approx \frac{\kappa}{2},
\end{align}
where we have  defined
\begin{equation}
    \Gamma_{eg}=C\frac{\kappa^2}{\kappa^2/4+ \delta^2} \widetilde{\Gamma}.
    \label{S.eq:Gamma_eg}
\end{equation}
 $C$ is the cooperativity, which quantifies the strength of the plasmon-molecule interaction relative to the system losses. 
Explicitly:
\begin{equation}
    C = \frac{\Lambda^2}{\kappa\widetilde{\Gamma}}.
    \label{S.eq:cooperativity}
\end{equation}

According to 
\Eqref{S.eq:spectrum_lorentzians}  the block \Eqref{S.eq:L2} will thus generate a double lorentzian peak structure in the emission spectrum with HWHM given by $\Gamma_\pm$ and peak maximum frequency $E_\pm$. 
The left $v_\pm$  and right $w_\pm$ by definition verify $\mathcal{L}_{2\times 2}(0,g) v_\pm= \lambda_\pm v_\pm$ and $ w_\pm^t\mathcal{L}_{2\times 2}(0,g)= \lambda_\pm w_\pm^t$.
The explicit analytical expressions read
\begin{align}
v_\pm & = \frac{1}{(\lambda_\pm-i\Delta+\tilde\Gamma/2)^2 - \Lambda^2}(i\Lambda, \lambda_\pm-i\Delta+\widetilde\Gamma/2)^t,
\label{S.eq:vpm}
\\
w_\pm & = (-i\Lambda, \lambda_\pm^*+i\Delta+\widetilde\Gamma/2)^t,
\label{S.eq:wpm}
\end{align}
Note that in this case, as  in Eqs.~ \eqref{S.eq:first_corre_final_A}, \eqref{S.eq:first_corre_final_B} and \eqref{S.eq:spectrum_lorentzians}, we always have the product between $(w_j, v_j)$. We can normalize one of them ensuring that $(w_i, v_j) = \delta_{ij}$.
We denote now explicitly the components of the vectors 
by an index $x$ or $y$:
$v_\pm =(v_{x,\pm}, v_{y,\pm})^t$ and $w_\pm =(w_{x,\pm}, w_{y,\pm})^t$. 
This allows to write the operators
$v_\pm$ and $w_\pm$ as:
\begin{align}
v_\pm& = v_{x,\pm}\ket{0,g}\bra{0,e}+v_{y,\pm}\ket{0,g}\bra{1,g} ,
\\
w_{\pm}& = w_{x, \pm}\ket{0,g}\bra{0,e}+w_{y,\pm}\ket{0,g}\bra{1,g}.
\end{align}
From \Eqref{S.eq:first_corre_final_A} we obtain the coefficients $A_\pm$,
\begin{equation}
A_\pm = v_{y, \pm} .
\label{S.eq:Apm}
\end{equation}

The second coefficient $B_\pm$ is obtained from \Eqref{S.eq:first_corre_final_B}:
\begin{equation}
B_\pm  = w^*_{x,\pm}\rho_{1g,0e}^{\text{st}}  + w^*_{y,\pm} P_{1g}^{\text{st}} .
\label{S.eq:Bpm}
\end{equation}
We remind that the only coherences that do not vanish in the steady state are $\rho_{1g,0e}^{\text{st}} $ and $\rho_{0e,1g}^{\text{st}} $ (see \Eqref{S.eq:block9x9}).
By defining defining $M_\pm = \rm ReA_\pm B_\pm$, and taking into account Eqs.~ \eqref{S.eq:vpm}, \eqref{S.eq:wpm} and \eqref{S.eq:Apm}, \eqref{S.eq:Bpm}.
\begin{equation}
    M_\pm = {\rm Re} \frac{P_{1g}^{\text{st}}  +i \Lambda/(\lambda_\pm-i\Delta+\widetilde\Gamma/2)\rho_{1g,0e}^{\text{st}} }{1 - [\Lambda/(\lambda_\pm-i\Delta+\widetilde\Gamma/2]^2}.
    \label{S.eq:Mpm}
\end{equation}
According to \Eqref{S.eq:spectrum_lorentzians} the spectrum is a sum of lorentzians that we define as:
\begin{equation}
    L_{x,y}(\omega) = \frac{1}{\pi}\frac{y}{(\omega-x)^2+y^2},
\end{equation}
where $x$ is the position and $y$ is HWHM. Thus we compile the results for the peak at $\omega_p$ from \Eqref{S.eq:peak_omegap}, the Lorentzian shape determined by the eigenvalues in \Eqref{S.eq:lambda.pm}, and the weight given by \Eqref{S.eq:Mpm}, obtaining the final expression for the spectrum:
\begin{equation}
    \frac{S(\omega)}{\kappa} = \sum_{q=0,d}P_{1q}^{\text{st}} L_{\omega_p,\kappa/2}(\omega) \ + \sum_{\mu=\pm}M_\mu  L_{E_\mu, \Gamma_\mu} (\omega).
\label{S.eq:spectrum_final}
\end{equation}
Each time the state $\ket{0,e}$ is populated, it interacts with $\ket{1,g}$ through the plasmon-molecule coupling strength $\Lambda$. From $\ket{1,g}$, the system can emit light at rates $\Gamma_\pm$, remove an excited electron at rate $\Gamma_{\alpha g}$, transitioning to $\ket{1,0}$, or add an electron at rate $\Gamma_{\alpha e}$, reaching $\ket{1,d}$. However, in the large cooperativity regime, we find that $\Gamma_{eg} \gg \Gamma_{\alpha\sigma}$ with $\sigma = g,e$. Consequently, we can neglect the Lorentzian contributions of the states $P_{1,q}$ for $q = 0,d$,
\begin{equation}
        \frac{S(\omega)}{\kappa} \approx \sum_{\mu=\pm}M_\mu L_{E_\mu, \Gamma_\mu} (\omega).
\label{S.eq:spectrum_final_c>1}
\end{equation}

\subsection{Second-order correlation function}
\label{S.sec:g2}
To characterize the statistical properties of the emitted light and assess photon correlations, we consider the second-order correlation function, defined as (see  Sec 3 of \cite{walls2008}), 
\begin{equation}
g^{(2)}(\tau) =\frac{\langle a^\dagger(0) a^\dagger (\tau) a(\tau) a(0) \rangle}{\langle a^\dagger a\rangle^2}.
\label{S.eq:second_order_correlation}
\end{equation}  
Here, $a^\dagger$ and $a$ are the creation and annihilation operators of the plasmonic mode, respectively, and $\tau$ denotes the time delay between photon detection events. This function provides insights into the nature of photon emission, distinguishing between classical and quantum light sources. The quantum light sources, such as single-photon emitters, display antibunching behavior characterized by $g^{(2)}(0) < 1 $, with an ideal single-photon source satisfying $g^{(2)}(0) = 0 $ .  

Using trace invariance, \Eqref{S.eq:second_order_correlation} can be rewritten in terms of the jump operator in \Eqref{S.eq:photonic_jump}.
\begin{equation}
\begin{aligned}
   g^{(2)}(\tau)  &=  \frac{\tr{a^\dagger(\tau)a(\tau)[a\rho^{\text{st}} a^\dagger]}}{\tr{a\rho^{\text{st}}  a^\dagger}}
   \\
    & =   \frac{\tr{\mathcal{J}_{\rm ph}(\tau)[\mathcal{J}_{\rm ph}\rho^{\text{st}} ]}}{\tr{\mathcal{J}_{\rm ph}\rho^{\text{st}} }}.
\end{aligned}
\end{equation}
By applying the quantum regression theorem and employing the left/right Liouvillian eigenvector expansion as we did in \Secref{S.sec:photoemission_spectrum}, we obtain the expression
\begin{equation}
\begin{aligned}
g^{(2)}(\tau) & = \frac{\sum_ie^{\lambda_i\tau}(w_0, \mathcal{J}_{\rm ph}v_i)(w_i, \mathcal{J}_{\rm ph}\underline{\rho}^{\text{st}} )}{|(w_0, \mathcal{J}_{\rm ph}\underline{\rho}^{\text{st}} )|^2}.
\label{S.eq:g2_left/right}
\end{aligned}
\end{equation}
\section{Analytical solutions above the first photon emission threshold (star-bias point)}
The following results are derived within the plateau defined by $\Delta < \mu_s - \varepsilon < U$ and $\mu_t - \varepsilon < 0$. A specific point ($\star$) within this plateau corresponds to the parameter set $(\mu_s, \mu_t) = \varepsilon + (1.4, -0.5)\omega_p$ with $\Delta = 0.7\omega_p$, $\varepsilon = -0.4\omega_p$, and $U = 2\omega_p$. For analytical convenience, we set the temperature to $k_B T = 10^{-2} \omega_p$, effectively suppressing thermal effects.

The transition rates in \eqref{S.eq:electronic_dissipator_rates} contain the tunneling rates $\Gamma_{\alpha\sigma}$. For simplicity, we will assume that $\Gamma_{\alpha g} = \Gamma_{\alpha e} = \Gamma_\alpha$.  We choose $\Gamma_s = 5 \Gamma_t = 5 \times 10^{-6} \omega_p$, where \( \alpha \) refers to either the tip $\Gamma_t$ or the substrate $\Gamma_s$.

We take $\kappa = 5\times10^{-2} \omega_p$ ensuring $\Gamma_\alpha \ll \kappa$.

\subsection{Stationary solution for the density matrix}
As explained in \Secref{S.sec:liouvillian_structure}, the only block in the Liouvillian that yields a nonzero density matrix in the stationary regime is the $9 \times 9$ block given in \Eqref{S.eq:block9x9}.
In this configuration, the doubly occupied state is inaccessible, allowing us to significantly reduce the Hilbert space dimension from a  $9 \times 9 $ matrix to a $ 7 \times 7$  matrix. $\underline{\rho}_7 = (P_{00}, P_{10}, P_{0g}, P_{0e}, P_{1g},\rho_{1g,0e}, \rho_{0e,1g})^T$.
When the doubly occupied states are discarded in \Eqref{S.eq:block9x9}, the transition rates given by \Eqref{S.eq:electronic_dissipator_rates} simplify to the following:
\begin{align}
    \Gamma_{0\rightarrow}  & = \Gamma_{0g} + \Gamma_{0e} = 2\Gamma_s,
    \\
    \Gamma_{0g} & = \Gamma_{t}f_t^+(\varepsilon)  + \Gamma_{s}f_s^+(\varepsilon)  = \Gamma_s,
    \label{S.eq:Gamma0g}
    \\
    \Gamma_{0e} & = \Gamma_{t}f_t^+(\varepsilon +\Delta)  + \Gamma_{s}f_s^+(\varepsilon +\Delta)  = \Gamma_s,
    \label{S.eq:Gamma0e}
     \\
    \Gamma_{g0} & = \Gamma_{t}f_t^-(-\varepsilon)  + \Gamma_{s}f_s^-(-\varepsilon)  = \Gamma_t,
    \label{S.eq:gamma_g0}
    \\
      \Gamma_{e0} & =\Gamma_{t}f_t^-(-\varepsilon-\Delta)  + \Gamma_{s}f_s^-(-\varepsilon-\Delta)  = \Gamma_t,
    \label{S.eq:gamma_e0}
\end{align}
where $f_\alpha^\pm(\epsilon) = [e^{\pm(\epsilon -\mu_\alpha)/k_B T} + 1]^{-1}$.
We can now insert these rates into \Eqref{S.eq:block9x9}, excluding the double-occupied states. In this context, we obtain the reduced $7\times7$ matrix, where the density matrix evolution is given by $\underline{\dot\rho}_{7\times 7} = \check{\mathcal{L}}_{7\star}\underline{\rho}_{7\times 7}$. Thus,
\begin{equation}
\check{\mathcal{L}}_{7\star}=
\begin{pmatrix}
-2\Gamma_s &   \kappa& \Gamma_t & \Gamma_t &  0 & 0 &0
\\
0& - \kappa &  0 & 0 & \Gamma_t & 0 &0
\\
\Gamma_s &  0&-\Gamma_t&0 & \kappa  & 0&0
\\
\Gamma_s & 0 & 0 &-\Gamma_t & 0 & -i\Lambda & i\Lambda
\\
0 & \Gamma_s &  0& 0 & - \kappa  &  i\Lambda & -i\Lambda
\\

0 & 0 & 0& -i\Lambda & i\Lambda  & z^* & 0
\\
0 & 0 & 0& i\Lambda & -i\Lambda  & 0&z 
\label{S.eq:block9x9_star}
\end{pmatrix} ,
\end{equation}

where $z = -\kappa/2 + i\delta$. The stationary solution of the density matrix is obtained by solving $\check{\mathcal{L}}_{7\star} \underline{\rho}_{7\times 7} = 0$, along with the probability conservation condition $P_{00} + P_{0g} + P_{0e} + P_{1g} = 1$, under the assumption that $P_{10} \ll P_{00}, P_{0g}, P_{0e}, P_{1g}$. The resulting expressions are:
\begin{align}
    P_{00}^{\text{st}}  & = \frac{\Gamma_t}{\Gamma_t+2\Gamma_s},
    \label{S.eq:P00_st_star}
    \\
    P_{10}^{\text{st}}  & = \frac{\Gamma_t\Gamma_s}{\kappa^2} 2\eta P_{00}^{\text{st}} ,
    \\
    P_{0g}^{\text{st}}  &= \frac{\Gamma_s}{\Gamma_t}\left(1 +  2\eta\right)P_{00}^{\text{st}} ,
    \label{S.eq:P0g_st_star}
    \\
    P_{0e}^{\text{st}}  &= \frac{\Gamma_s}{\kappa}\left(\frac{\Gamma_{eg} +\kappa}{\Gamma_{eg}}\right)2\eta P_{00}^{\text{st}} ,
    \label{S.eq:P0e_st_star}
    \\
    P_{1g}^{\text{st}} &= \frac{\Gamma_s}{\kappa}2\eta P_{00}^{\text{st}} ,
        \label{S.eq:P1g_st}
    \\
    \rho_{1g, 0e}^{\text{st}}  & = i(i\delta -\kappa/2)\frac{\Gamma_s}{\kappa\Lambda} 2\eta P_{00}^{\text{st}} .
 \label{S.eq:rho_1g_0e_st}
\end{align}
where $\Gamma_{eg} = \kappa \Lambda^2/(\kappa^2/4 + \delta^2)$ and $2\eta =\Gamma_{eg}/(\Gamma_t + \Gamma_{eg})$ with $\eta$ the quantum yield calculated in \Eqref{S.eq:eta_star}. The fact of appearing $\Gamma_t$ is because of the specific choice that we have made for the $\star$ bias conditions. 
If the bias polarity is reversed the expression still holds with the indices  $s$ and $t$ exchanged.

\subsection{Quantum yield}

In this section, we analytically compute the electronic current. Since the current satisfies the condition that its flux at the tip electrode is equal and opposite to the flux at the substrate, i.e., $I_t = -I_s$, we focus on calculating the current at the substrate $I_s$, which is significantly simpler. To achieve this, we utilize Eqs.~\eqref{S.eq:electronic_current}, \eqref{S.eq:Gamma0g} and \eqref{S.eq:Gamma0e},
\begin{equation}
    I_{s} = 2\Gamma_s P_{00}^{\text{st}} .
\end{equation}
Then we can now calculate the photo current according to \Eqref{S.eq:photonic_current},
\begin{equation}
    I_{\rm ph} = \kappa (P_{10}^{\text{st}}  + P_{1g}^{\text{st}} ) \approx \kappa  P_{1g}^{\text{st}} .
\end{equation}
We then define the quantum yield ($\eta$) as the ratio of the photon flux to the electron flux, i.e.,
\begin{equation}
    \eta = \frac{I_{\rm ph}}{I_{s} }  = \frac{\kappa P_{1g}}{2\Gamma_s P_{00}^{\text{st}} } = \frac{1}{2}\frac{\Gamma_{eg}}{ \Gamma_t + \Gamma_{eg}}.
    \label{S.eq:eta_star}
\end{equation}
where we have inserted Eqs.~\eqref{S.eq:P1g_st} and \eqref{S.eq:P00_st_star}.

\subsection{Photon emission spectrum}
\label{S.sec:spectrum_analytics}
In this section, we present the emission spectrum given in \Eqref{S.eq:spectrum_final} for the bias voltage configuration under consideration. To begin, we express the eigenvalues $\lambda_\pm$ from \Eqref{S.eq:lambda.pm}, 
 \begin{align}
     \lambda_\pm =  i\frac{(\omega_p + \Delta)}{2} - \Gamma_t-\frac{\kappa}{4}\pm\frac{1}{2}\sqrt{(\kappa/2-i\delta)^2-4\Lambda^2}.
 \end{align}
where we have taken into account the explicit value of the total decay rate as $\widetilde\Gamma=\Gamma_{g\rightarrow}+\Gamma_{e\rightarrow}=2\Gamma_t$ and $\Gamma_{g\rightarrow}=\Gamma_t$ according to Eqs.~ \eqref{S.eq:gamma_g0} and \eqref{S.eq:gamma_e0}. The weight $M_\pm$ in \eqref{S.eq:Mpm} can be rewrite including the value of the populations we find in Eqs \eqref{S.eq:P1g_st}  and \eqref{S.eq:rho_1g_0e_st} as $M_\pm = P_{1g}^{\text{st}}W_\pm$ where,
\begin{equation}
    W_\pm = {\rm Re} \frac{\Lambda^2}{\Lambda^2- (\lambda_\pm-i\Delta + \Gamma_t)^2}.
\end{equation}
Finally we can easily express the spectrum:
\begin{equation}
    \frac{S(\omega)}{2\eta\Gamma_sP_{00}^{\text{st}}} = \frac{\Gamma_t}{\kappa}L_{\omega_p,\kappa/2}(\omega) + \sum_{\mu=\pm} W_\mu L_{E_\mu, \Gamma_\mu} (\omega),
\end{equation}
which is the expression provided in the main text.

\section{Electronic and photonic currents in different coupling regimes}
\label{S.sec:electronic_photonic_current_description}
For completeness, we provide a detailed analysis of the electronic and photonic currents for different bias voltage configurations. This extends Fig.~1(b,c) in the main text, where we analyze the behavior in the regime of plasmon-molecule coupling strength.

\begin{figure}
    \includegraphics[width=\columnwidth]{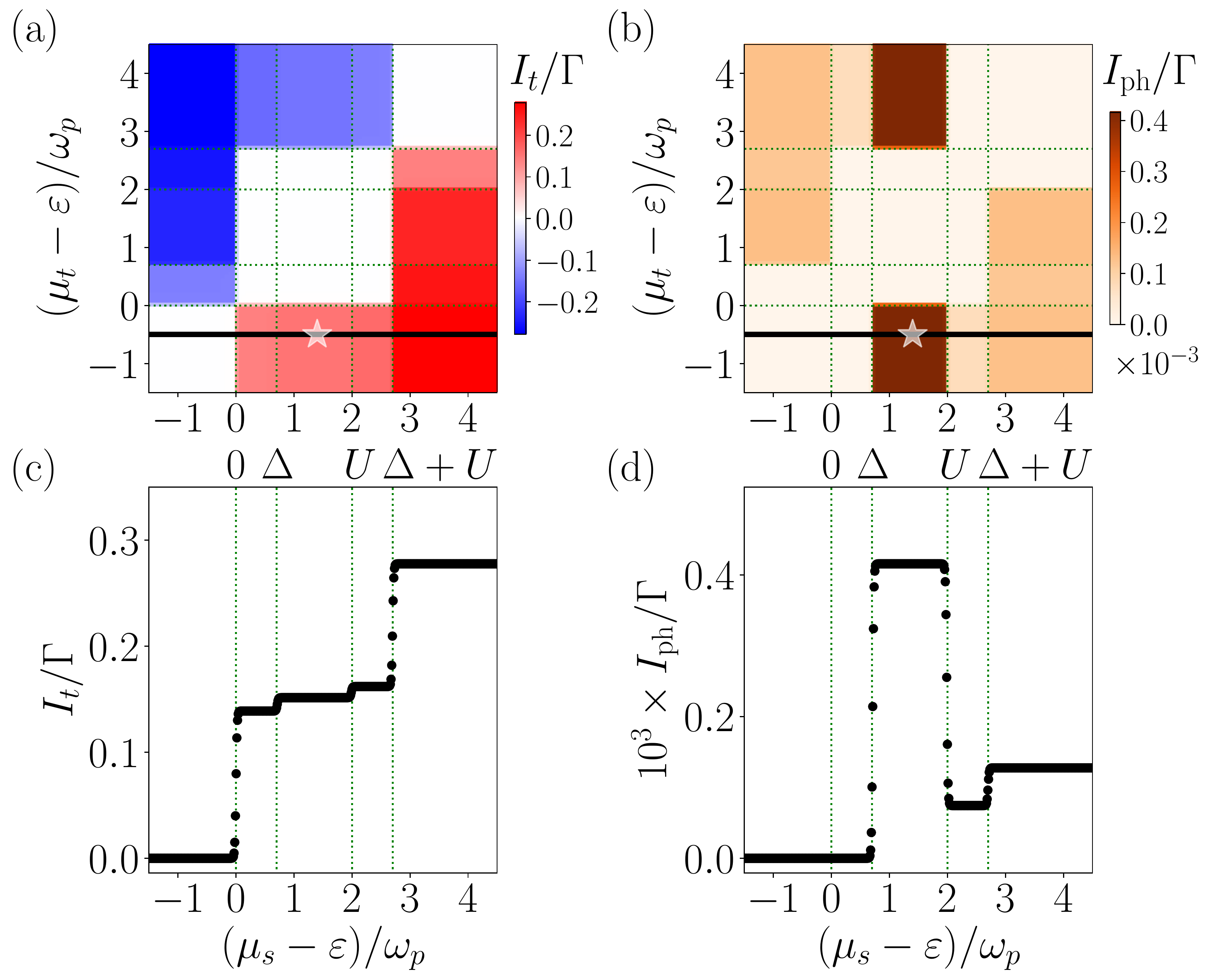}
    \caption{(a,b) Current profiles as a function of $\mu_s$ for a fixed $\mu_t - \varepsilon = -0.5\omega_p$ (black line in panels b-c), illustrating distinct steps corresponding to the activation of various electron injection processes.  
    The calculations were performed for $\Lambda/\kappa = 0.002$, $\Delta = 0.7\omega_p$, $\varepsilon = -0.4\omega_p$, $U = 2\omega_p$, $\kappa =  0.05\omega_p$, $k_B T =0.01\omega_p$ and $\Gamma_s = 5\Gamma_t = 5\times 10^{-6} \omega_p$.}
    \label{S.fig:current.I}
\end{figure}

We begin our discussion with the weak coupling regime, $\Lambda = 0.02\kappa$, as shown in \Figref{S.fig:current.I}(c) presents a cross-section at $\mu_t - \varepsilon = -0.5$, where four distinct steps can be observed. The first step occurs at $\mu_s - \varepsilon = 0$, corresponding to the energy required to remove an electron from the ground state. The second step appears at $\mu_s - \varepsilon = \Delta$, which is the energy necessary to add an electron to the excited state. The third step emerges at $\mu_s - \varepsilon = U$, representing the transition from $\ket{0, e}$ to $\ket{0, d}$. Finally, the last step corresponds to the transition $\ket{0, g} \rightarrow \ket{0, d}$.

Regarding the photocurrent in \Figref{S.fig:current.I}(c), we observe that photon emission occurs at the second step of the electronic current, specifically at $\mu_s - \varepsilon = \Delta$. This can be understood as follows: at this energy, the excited state $\ket{0, e}$ becomes populated, which subsequently interacts with $\ket{1, g}$ through the coupling strength $\Lambda$, leading to photon emission. Beyond this point, a dip in the photocurrent appears at $\mu_s - \varepsilon = U$. The reason for this suppression is that, at this energy, the system has two competing pathways: it can either transition to $\ket{0, d}$ or continue interacting with $\ket{1, g}$. The final step in the photocurrent is associated with the transition $\ket{0, g} \rightarrow \ket{0, d}$. Once the system reaches $\ket{0, d}$, it can decay to $\ket{0, e}$ by emitting an electron, leading to an overall increase in the total flux.

\begin{figure}
    \includegraphics[width=\columnwidth]{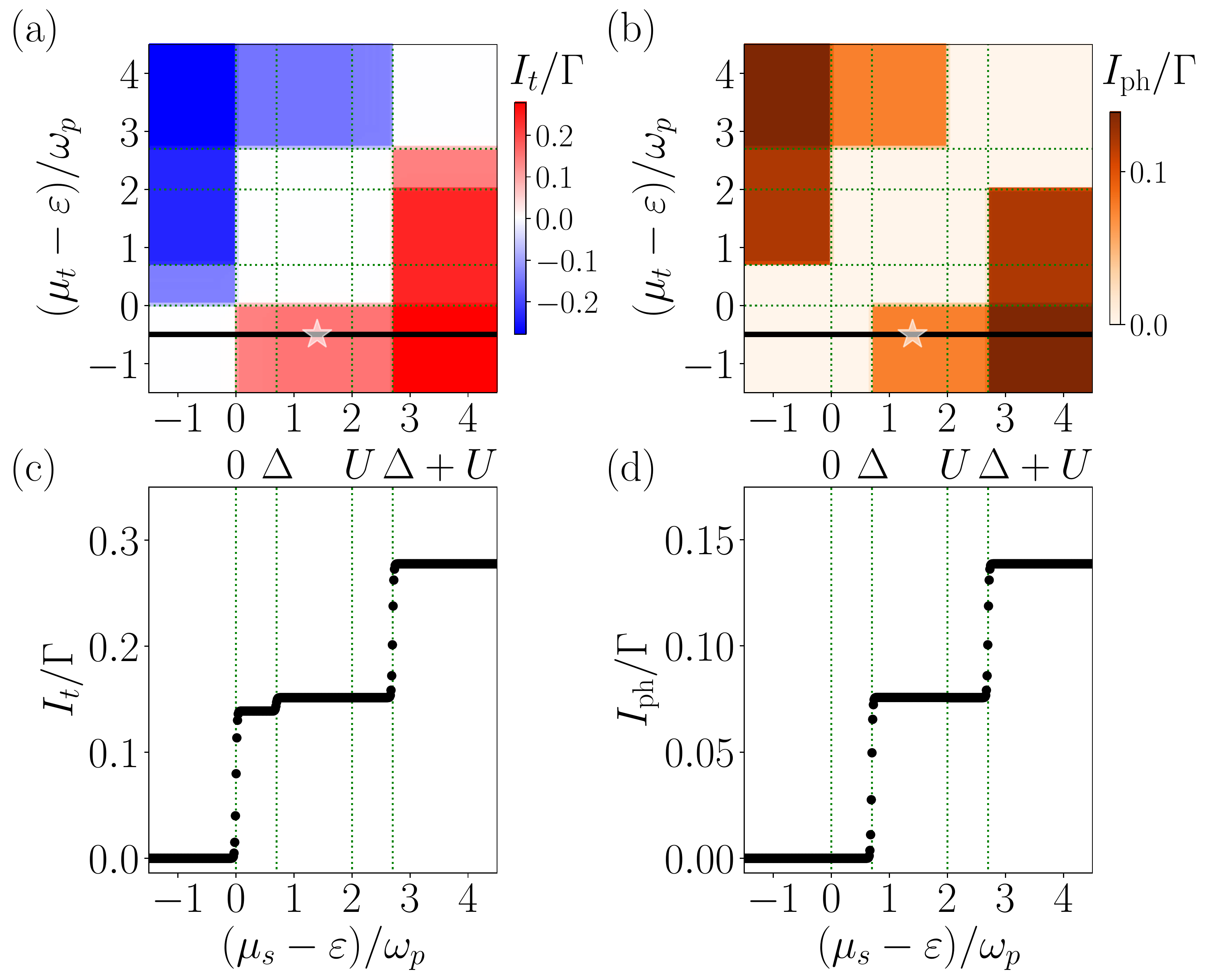}
    \caption{As \Figref{S.fig:current.I}, but in the strong-coupling regime $\Lambda/\kappa=1.6$.}
    \label{S.fig:current.III}
\end{figure}

In the strong coupling regime, as shown in \Figref{S.fig:current.III}, the only noticeable difference compared to the weak coupling case (\Figref{S.fig:current.I}) is the suppression of the step at $\mu_s - \varepsilon = U$. As just explained, at this energy, the system in state $\ket{0,e}$ has two possible transitions: either adding an electron to the ground state or interacting with $\ket{1,g}$ via the coupling strength $\Lambda$. However, in the strong coupling regime, each time the system reaches $\ket{0,e}$, it rapidly interacts with $\ket{1,g}$, effectively reducing the probability of the transition $\ket{0, e} \rightarrow \ket{0, d}$. Consequently, the step associated with this transition is suppressed.

\section{Effective model with structured bath}
\label{S.sec:structured}
The model presented in the main text captures the transition from the weak to the strong coupling regime. 
In this section, we develop an effective model specifically suited for the weak coupling regime, where the plasmonic mode is described as a structured environment.
We construct this model by first considering a single photonic mode coupled to a continuum of modes with a constant density of states, as described by the Hamiltonian for the problem at hand.
Next, we calculate the tunneling rates for the electrons, assuming that the final states correspond to the eigenstates obtained through the previous procedure.
While this approach does not allow for the study of the time evolution of the photonic modes, it is simpler to implement and effectively describes the weak coupling limit.
Moreover, the explicit derivation from the Hamiltonian used in this approach establishes a connection between the full master equation approach and the simplified structured environment approach.

\subsection{Hamiltonian}

We begin writing the Hamiltonian 
defined in \Eqref{S.eq:Hamiltonian} as  
\begin{equation}  
\op{H} = \op{H}_{\rm S'} + \op{H}_{\rm B'} + \op{H}_{\rm S'B'}
,
\label{S.eq:H_full_sb}
\end{equation}  
where we reshuffled the terms of the original Hamiltonian in a different way. 
We will regard the central system $S'$ as the purely electronic part, 
{\em i.e.}
\begin{equation}
    \op{H}_{\rm S'} = \op{H}_{\rm m},
\label{S.eq:Hm_sb}
\end{equation}
where $\op{H}_{\rm m}$ is given in \eqref{S.eq:Hm}. 
Next, we incorporate the plasmonic mode $\op{H}_{\rm p}$  from \Eqref{S.eq:Hp}, along with its interaction with the free electromagnetic modes $\op{H}_{\rm p-em}$ from \Eqref{S.eq:Hpem} into the new bath Hamiltonian.  
\begin{equation}  
\op{H}_{\rm B'} = \op{H}_{\rm leads} + \op{H}_{\rm sb},
\end{equation}
where $\op{H}_{\rm leads}$ is defined in Eqs.~\eqref{S.eq:Hleads}
and we defined what we call the structured bath Hamiltonian,
\begin{equation}
\op{H}_{\rm sb} = \op{H}_{\rm em} + \op{H}_{\rm p} + \op{H}_{\rm p-em},
\label{S.eq:Hsb}
\end{equation} 
where $\op{H}_{\rm em}$ is defined by \eqref{S.eq:Hem}.
The structured bath term $\op{H}_{\rm sb}$ will be the focus of next section.
Finally, the system-bath interaction term is:
\begin{equation}
    \op{H}_{\rm S'B'} = \op{H}_{\rm tun} + \op{V},
    \label{S:eq.HSB_eff}
\end{equation}
where the plasmon-molecule interaction $\op{V}$ is given in \eqref{S.eq:V}.

\subsection{Diagonalization of the structured bath Hamiltonian}
To begin our analysis, we express the Hamiltonian describing the structured bath, as given in \Eqref{S.eq:Hsb}. Specifically, the interaction between the plasmonic mode and the set of electromagnetic modes, $H_{\rm p-em}$, as defined in \Eqref{S.eq:Hpem}, is assumed to have a uniform coupling strength across all terms: $g = g_\nu$.
This gives:
\begin{equation}
\begin{aligned}
\op{H}_{\rm sb} = &~ \omega_p \opd{a}\op{a} +\sum_{\nu=1}^N\omega_\nu\opd{b}_\nu\op{b}_\nu 
\\
&~+ g\left(\opd{a}\sum_{\nu=1}^N\op{b}_\nu+\op{a}\sum_{\nu=1}^N\opd{b}_\nu\right).
\label{S.eq:Hsb_explicit}
\end{aligned}
\end{equation}

We now diagonalize the $H_{\rm sb}$ following standard methods (procedure outlined in Section $C_I$ of \cite{cohen-tannoudji_atom_1998})
$H_{\rm sb}$ can be written as a quadratic form:
\begin{equation}
\op{H}_{\rm sb}=\opd{\mathbf{b}}\mathbf{H}_{\rm sb}\op{\mathbf{b}},
\end{equation}
where we introduced the vector of operators:
\begin{equation}
\op{\mathbf{b}} = (\op{a}, \op{b}_1,\cdots, \op{b}_N )^T,
\end{equation}
and the matrix elements of the structure bath Hamiltonian $\op{H}_{\rm sb}$:
\begin{equation}
\mathbf{H}_{\rm sb}=
\begin{pmatrix}
\omega_p & g & g &\cdots & g\\
g & \omega_1 & 0 & \cdots & 0 \\
g & 0 & \omega_2 & \cdots & 0 \\
\vdots & \vdots & \vdots &\ddots & \vdots\\
g & 0 & 0 & \cdots & \omega_N
\end{pmatrix}.
\label{S.eq:Hsb_matrix}
\end{equation}
If we diagonalize this matrix, we can construct a transformation matrix $\mathbf{U}$ whose columns correspond to the eigenvectors of $\mathbf{H}$. Denoting the diagonalized matrix as $\mathbf{\tilde{H}}$, we obtain the relation:
\begin{equation}
\mathbf{H}_{\rm sb} = \mathbf{U}\mathbf{\tilde{H}}_{\rm sb}\mathbf{U}^{-1}.
\end{equation}
Then we can relate the operators as:
\begin{equation}
\op{\mathbf{b}}=\mathbf{U}\tilde{\mathbf{b}}.
\label{S.eq:ubbtilde}
\end{equation}
The structured bath Hamiltonian, once diagonalized ($\op{H}_{\rm sb}$), spans an $(N+1)$-dimensional space.
\begin{equation}
\tilde{H}_{\rm sb} = \sum_{\mu=1}^{N+1}\omega_\mu\tilde{b}_\mu^\dagger \tilde{b}_\mu.
\label{S.eq:Hsb_diagonal}
\end{equation}
Our objective is to determine the eigenenergies $\omega_\mu$ and the modes $\tilde{b}_\mu$ in terms of the system parameters. This problem can be analytically solved. The diagonal matrix elements of \Eqref{S.eq:Hsb_matrix} are given by:
\begin{align}
\bra{\phi}\op{H}&_{p} \ket{\phi} = \omega_p,
\\
\bra{\nu}\op{H}&_{\rm em} \ket{\nu} = \omega_\nu,
\end{align}
where $\ket{\phi}$ is an eigenstate of $\op{H}_{p}$ with energy $\omega_p$, and $\ket{\nu}$ is an eigenstate of $\op{H}_{\rm em}$ with energy $\omega_\nu$. Consequently, the off-diagonal elements take the form:
\begin{align}
\bra{\phi}\op{H}_{\rm sb}\ket{\nu}   & = g,  
\\
\bra{\phi}\op{H}_{\rm sb}\ket{\phi}  & = 0, 
\\
\bra{\nu}\op{H}_{\rm sb}\ket{\nu}   &= 0. 
\end{align}
In this basis, the completeness relation is given by
\begin{equation}
\ket{\phi}\bra{\phi}+\sum_{\nu=1}^N\ket{\nu}\bra{\nu} = \mathbb{1}.
\label{S.eq:norm_phi_p}
\end{equation}
We now assume that the eigenstates of the total Hamiltonian $\op{H}_{\rm sb}$ are given by
\begin{equation}
\op{H}_{\rm sb}\ket{\mu} =\omega_\mu\ket{\mu}.
\end{equation}
Following \cite{cohen-tannoudji_atom_1998} we obtain the energy spectrum of the new modes $\tilde{b}_\mu$ and the coefficients:
\begin{align}
\omega_\mu  & = \omega_p+ \frac{\kappa}{\tan\frac{\pi\omega_\mu}{\Delta_o}}\label{S.eq:Hbs_omega_mu},
\\
\braket{\phi|\mu} & = \frac{g}{\sqrt{g^2 + \kappa^2/4 + (\omega_\mu-\omega_p)^2}}  ,
\label{S.eq:phimu_1}
\\
\braket{\nu|\mu} &= \frac{g^2/(\omega_\mu-\omega_\nu)}{\sqrt{g^2 + \kappa^2/4 + (\omega_\mu-\omega_p)^2}} .
\label{S.eq:numu}
\end{align}
To derive the previous equations, we assumed that the spectrum of the free electromagnetic modes in $H_{\rm em}$ is uniformely spaced, so that $\omega_\nu = \nu\Delta_o$, where $\Delta_o$ represents the mode energy spacing.
The quantity $\kappa=2\pi g^2/\Delta_o$ plays the role of a damping of the discrete mode and has simple interpretation in terms of Fermi's golden rule, since $1/\Delta_o$ is density of states.
In order to impose the description 
of a damped mode one can assume that 
for $\Delta_o \to 0$, the ratio $g^2/\Delta_o$ remains constant and equal to $\kappa/2\pi$.

The coefficients in Eqs.~\eqref{S.eq:phimu_1} and \eqref{S.eq:numu}  are the elements of the transformation matrix $\mathbf{U}$ in \Eqref{S.eq:ubbtilde} and allow us to relate the operators in the following way:
\begin{equation}
\op{a} = \sum_\mu \braket{\phi|\mu}  \tilde{b}_\mu
,
\quad
    \op{b}_ \nu= \sum_\mu \braket{\nu|\mu}  \tilde{b}_\mu.
    \label{S.eq:phimu}
\end{equation}

\subsection{Calculation of the decay rate to the structured environment}

In this section, we calculate the transition rates induced by the system-bath interaction as described in \Eqref{S:eq.HSB_eff} 
in terms of the eigenstates of 
$H_{sb}$.
The plasmon-molecule interaction is expressed in terms of the new operators $\tilde{b}_\mu$ through Eqs. \eqref{S.eq:phimu}:
\begin{equation}
\op{V}  = \op{d}_g^\dagger\op{d}_e\sum_\mu \Lambda_\mu\tilde{b}_\mu^\dagger+ \op{d}_e^\dagger\op{d}_g\sum_\mu  \Lambda_\mu^*\tilde{b}_\mu ,
\end{equation}
where 
$\Lambda_\mu = \Lambda \braket{\mu|\phi}$. 
We now calculate using Fermi's golden rule the transition rate from the  state $\ket{0}_{\rm sb}\otimes \ket{e}$ 
to the state $\tilde b_\mu^\dag\ket{0}_{\rm sb}\otimes \ket{g}$,
with respective energy, 
$\varepsilon +\Delta$ and $\varepsilon + \omega_\mu$.
Here $|0\rangle_{sb}$ is the vacuum of the structured bath Hilbert space.
We obtain
\begin{align}
\Gamma_{eg} & =  2\pi\sum_\mu\left| \Lambda_\mu\right|^2\delta(\Delta - \omega_\mu)
\\
&=2\pi  \int d\omega \left(\frac{dN_\mu}{d\omega}\right)|\Lambda(\omega)|^2\delta(\Delta - \omega).
\end{align}
The density of states of the modes $\tilde{b}_\mu$ is approximately the same as that of the modes $\op{b}_\nu$,
since each interacting eigenvalue is found between two non-interacting eigenvalues.
As mentioned in the previous section, we have assumed that the density of states for the free electromagnetic modes is $1/\Delta_o$.
Therefore, we can write:
\begin{equation}
    \frac{dN_\mu}{d\omega} \approx  \frac{dN_\nu}{d\omega} = \frac{1}{\Delta_o}
    .
\end{equation}
We can define now the spectral density of states $J(\omega)$ of the structured bath as follows:
\begin{align}
    J(\omega) & = \left(\frac{dN_\mu}{d\omega}\right)|\Lambda(\omega)|^2
    \\
    & = \frac{1}{2\pi}\frac{\Lambda^2\kappa}{\kappa^2/4+(\omega-\omega_p)^2}
    \label{S:eq.spectral_sb}
    .
\end{align}
Where have taken $\Delta_o\rightarrow 0 $ and $g^2/\Delta_0= \kappa/2\pi$.
Finally we obtain for the transition rate:
\begin{equation}
\Gamma_{eg} =\frac{\Lambda^2 \kappa}{\kappa^2/4+\delta^2}.
\end{equation}
This expression matches the expected form, as derived in \Eqref{S.eq:Gamma_eg}, and characterizes a mode with finite width resulting from its coupling to the environment.

\subsection{Master equation for the electronic system}
\label{S.sec:master_equation_sb}
Although structured photonic environments generally induce non-Markovian dynamics due to their non-flat spectral density and long-lived correlations (see Sec. 10.1.2 in \cite{petruccione2002}), in the limit $\Gamma_{\alpha\sigma}\!\ll\!\kappa$ the situation simplifies considerably. Specifically, the correlation time of the bath is given by $\tau_{\text{mem}} \sim 1/\kappa$, which is much shorter than the inverse of the system's excitation rate, $\Gamma_{\alpha\sigma}^{-1}$. This separation of timescales allows us to work within the Markovian approximation, effectively neglecting memory effects and non-local temporal correlations. Such treatment remains valid as long as the system–bath coupling remains weak and the bath correlations decay rapidly compared to the system dynamics. It is important to note that this assumption no longer holds when probing dynamics on timescales $\tau \sim 1/\kappa$, such as those associated with Rabi oscillations.
Within this approximation, we derive the master equation for the central system $S'$, consisting of the electronic degrees of freedom. Tracing out the fermionic and photonic environments leads to:
\begin{equation}
\partial_t\op{\rho} = -i[\op{H}_{m},\op{\rho}] + [\mathcal{L}_e^++\mathcal{L}_e^- + \mathcal{L}_{\rm sb}]\rho.
\label{S.eq:me_sb}
\end{equation}
where $\mathcal{L}_e^\pm$ are the tunneling terms derived in \Eqref{S.eq:electronic_dissipator_pm}, while 
\begin{equation}
\mathcal{L}_{\rm sb} = \Gamma_{eg}\left(\tilde{\sigma}\op{\rho}\tilde{\sigma}^\dagger - \{\tilde{\sigma}^\dagger\tilde{\sigma},\op{\rho}\}/2 \right),
\label{eq:L_rmodes}
\end{equation}
where the effective raising/lowering operators are defined as:
\begin{align}
\tilde{\sigma}&\equiv \op{d}_g^\dagger \op{d}_e = \ketbra{g}{e} ,
\\
\tilde{\sigma}&^\dagger\equiv\op{d}_e^\dagger\op{d}_g  =\ketbra{e}{g} .
\end{align}
Within the secular approximation, we know that the Liouvillian can be diagonalized in block form, where the single large block involves only the populations, since the coherences enter one-dimensional blocks.
\subsection{Analytical solution of $g^{(2)}(\tau)$}
We consider the second-order correlation function, 
\begin{equation}
    g^{(2)}(\tau) = \frac{\langle\tilde\sigma^\dagger(0)\tilde\sigma^\dagger(\tau)\tilde\sigma(\tau)\tilde\sigma(0)\rangle}{|\langle\tilde\sigma^\dagger\tilde\sigma\rangle|^2},
\end{equation}
which is a well-established quantity for characterizing the quantum statistical properties of emitted light. This expression, commonly used in resonance fluorescence and quantum optics, provides direct insight into photon correlations and the underlying emission dynamics of driven two-level systems  (See Sec. 10 of \cite{scully1997}).
We now apply the quantum regression theorem, following the same approach used in \Secref{S.sec:emission_spectrum},
\begin{align}
\langle\tilde\sigma^\dagger(0)\tilde\sigma^\dagger(\tau)\tilde\sigma(\tau)\tilde\sigma(0)\rangle=&~\tr{\tilde{\sigma}^\dagger\tilde{\sigma}e^{\mathcal{L}\tau}[\tilde{\sigma}\op{\rho}^{\text{st}} \tilde{\sigma}^\dagger]}
\\
= &~ P_e^{\text{st}} \bra{e}e^{\mathcal{L}\tau}[\ketbra{g}{g}]\ket{e}
\\
= &~
P_e^{\text{st}}  P_{e}(\tau)\bigg|_{\op{\rho}(0)=\ketbra{g}{g}}.
\end{align}
The second term above is the conditional probability of finding the system in the state $\ket{e}$ at time $\tau$, given that the system was in the state $\ket{g}$ at time $\tau=0$.
Indeed,  $\rho(\tau) = e^{\mathcal{L}\tau} \rho(0)$, if  $\rho(0) = \ketbra{g}{g}$ we get $\bra{e} \rho(\tau)\ket{e}  =\bra{e} e^{\mathcal{L}\tau}[\ketbra{g}{g}]\ket{e}$.
On the other hand,   $|\langle\tilde\sigma^\dagger\tilde\sigma\rangle|^2=(P_e)^2$. 
Thus:
\begin{equation}
g^{(2)}(\tau) =
\frac{P_e(\tau)\vert_{\op{\rho}(0)=\ketbra{g}{g}}}{P_e^{\text{st}} }
.
\label{S.eq:g2_sb}
\end{equation}
As stated in \Secref{S.sec:master_equation_sb}, the only large block is the one that governs the population dynamics, while each coherence term evolves independently. 
Therefore, we can restrict our analysis to the block defined by the populations.
By defining  $\underline{P} = (P_0, P_g, P_e, P_{d})^T$ we write a matrix expression for populations $\dot{\underline{P}}= \check{\mathcal{L}}\underline{P}$. 
The solution of this equation can be readily written in terms of the right eigenvectors of the Liouvillian operator: 
$\dot{\underline{P}}(\tau) =  c_0\underline{P}^{\text{st}}  + 
 \sum_{j=1,2, 3}c_je^{\lambda_j \tau}v_j$
where $(c_0, c_1, c_2, c_3)$ are determined by the initial conditions which in this case gives $\mathbf{P}(0)=(0, 1, 0, 0)^T$.
The $v_j$ are the right eigenvectors corresponding to the eigenvalues $\lambda_j$ of the Liouvillian $\check{\mathcal{L}}$. 
As discussed in \Secref{S.sec:emission_spectrum}, the right eigenvectors $v_j$ must be normalized such that $(w_i, v_j) = \delta_{ij}$, where $w_i$ is the left eigenvector of the Liouvillian. 
However, this normalization allows for some freedom in the choice of the normalization since the left and right eigenvectors are different. 
It is convenient to choose $v_0$ normalized, so that it represents the stationary distributions, for the other vectors any choice that guarantees $(w_i, v_j) = \delta_{ij}$ is sufficient.

The Liouvillian of this block can be
derived from \Eqref{S.eq:me_sb}. 
Below, we present its explicit expression for the bias voltage configuration indicated by a $\star$ in Figs. \eqref{S.fig:current.I}, and \eqref{S.fig:current.III}:
\begin{equation}
\check{\mathcal{L}}=
\begin{pmatrix}
-2\Gamma_s &  \Gamma_t & \Gamma_t
\\
\Gamma_s & -\Gamma_t &   \Gamma_{eg}
\\
\Gamma_s & 0 & -\Gamma_t -  \Gamma_{eg} 
\end{pmatrix}
.
\end{equation}
The eigenvalues read $\lambda_0 = 0 $, $\lambda_1 = -2\Gamma_s -\Gamma_t$ and $\lambda_2 = -\Gamma_t -  \Gamma_{eg}$, with corresponding eigenvectors whose normalization has ben choose such that $P(0) = (0, 1, 0)$:
\begin{align}
  \mathbf{v}_0&~ = \left(P_0^{\text{st}} , P_g^{\text{st}} , P_e^{\text{st}} \right)^T ,
\label{S.eq:populations_sb}
\\
\mathbf{v}_1&~= \frac{P_0^{\text{st}} }{2\Gamma_s -  \Gamma_{eg}}\left(  \Gamma_{eg}-2\Gamma_s,\Gamma_s-   \Gamma_{eg}, \Gamma_s\right)^T ,
\\
\mathbf{v}_2&~= \left(P_e^{\text{st}} + \frac{P_0^{\text{st}} }{2\Gamma_s-  \Gamma_{eg}}\right)\left(0, 1, -1\right)^T,
\end{align}
 where the stationary populations are,
\begin{align}
     P_0^{\text{\text{st}}} &=  \frac{\Gamma_t}{\Gamma_t+2\Gamma_s},
     \label{S.eq:P0_sb_st}
     \\
    P_g^{\text{\text{st}} } &= \frac{\Gamma_s}{\Gamma_t} \left(1 +\frac{\Gamma_{eg}}{\Gamma_{eg}+\Gamma_t}\right) P_0^{\text{\text{st}} },
    \label{S.eq:Pg_sb_st}
    \\
    P_e^{\text{\text{st}} } &= \frac{\Gamma_s}{\Gamma_t + \Gamma_{eg}} P_0^{\text{\text{st}}}.   
    \label{S.eq:Pe_sb_st}
\end{align}

It is interesting to see that these populations match, in the weak coupling limit, with those obtained in the full approach given by Eqs.~\eqref{S.eq:P00_st_star}, \eqref{S.eq:P0g_st_star}, and \eqref{S.eq:P0e_st_star}.

Then this gives:
\begin{equation}
\begin{aligned}
P_e\Big|_{\op{\rho}(0)=\ketbra{g}{g}} =&~ P_e^{\text{st}}  + \frac{\Gamma_s}{2\Gamma_s-  \Gamma_{eg}} P_0^{\text{st}} e^{-(2\Gamma_s+\Gamma_t)\tau} 
\\
&-\left(P_e^{\text{st}} + \frac{P_0^{\text{st}} }{2\Gamma_s-  \Gamma_{eg}}\right)e^{-(\Gamma_t+  \Gamma_{eg})\tau} .
\label{S.eq:Pe_evol_sb}
\end{aligned}
\end{equation}
We express the second-order correlation function in \Eqref{S.eq:g2_sb} using the populations from Eqs.~\eqref{S.eq:P0_sb_st}, \eqref{S.eq:Pe_sb_st} and \eqref{S.eq:Pe_evol_sb},
\begin{equation}
g^{(2)}(\tau) = 1 + Me^{-(2\Gamma_s+\Gamma_t)\tau} 
- (1+M)e^{-(\Gamma_t+  \Gamma_{eg})\tau},
\label{eq:g2_analytics_csmall}
\end{equation}
where $M=(\Gamma_t +   \Gamma_{eg})(2\Gamma_s -  \Gamma_{eg})$. While the analytical form clearly indicates a sum of two exponential terms in the correlation function, its expansion at short times surprisingly exhibits a quadratic time dependence,
\begin{equation}
    g^{(2)}(\tau) \approx
     (2\Gamma_s+\Gamma_t)(\Gamma_t+\Gamma_{eg})\tau^2/2
     .
\end{equation}
We refer to the main text for the discussion on the time dependence of $g^{(2)}(\tau)$.
\bibliography{bibliography}

\end{document}